\documentclass[compsoc,conference,a4paper,10pt,times]{IEEEtran}
\IEEEoverridecommandlockouts
\usepackage{tikz}
\usepackage{amsmath}

\usepackage{xcolor,color,xspace,enumerate,centernot,multirow,float,graphicx,
xcolor,caption,subcaption,textcomp,pgfplots,pgf-pie,tikz,listings,enumitem,
comment,adjustbox,mdframed,changepage,algorithm,algorithmic,soul,msc,tablefootnote}
\usepackage{filecontents}
\usepackage{xurl} % for usenix
\usepackage{makecell}
\usepackage{comment}

\usepackage[utf8]{inputenc}
\usetikzlibrary{shapes}
\usetikzlibrary{arrows.meta}
\usetikzlibrary{arrows}
\usetikzlibrary{positioning}
\usetikzlibrary{shadows}
\usetikzlibrary{backgrounds}
\usepackage[most]{tcolorbox}
\definecolor{lightblue}{rgb}{0.68, 0.85, 0.9}
\definecolor{lightpink}{rgb}{1.0, 0.71, 0.76}
  {\end{adjustwidth}}

\definecolor{darkgray}{gray}{0.6}

\UseRawInputEncoding

\usepackage{cite}
\usepackage{amsmath,amssymb,amsfonts}
\usepackage{algorithmic}
\usepackage{graphicx}
\usepackage{textcomp}
\usepackage{bmpsize}
\usepackage{xcolor}
\usepackage{lipsum}
\usepackage{xurl}
\usepackage[colorlinks=true,urlcolor=black]{hyperref}
\def\BibTeX{{\rm B\kern-.05em{\sc i\kern-.025em b}\kern-.08em
    T\kern-.1667em\lower.7ex\hbox{E}\kern-.125emX}}

\newcommand{\datasetA}{Dataset1 }
\newcommand{\datasetB}{Dataset2 }

\newcommand{\datasetANS}{Dataset1}
\newcommand{\datasetBNS}{Dataset2}

\newcommand{\datasetAsnark}{Dataset1\_SNARK }
\newcommand{\datasetBsnark}{Dataset2\_SNARK }

\newcommand{\datasetAsnarkNS}{Dataset1\_SNARK}
\newcommand{\datasetBsnarkNS}{Dataset2\_SNARK}

\newcommand{\liboqs}{\textbf{\texttt{liboqs}}}
\newcommand{\circl}{\textbf{\texttt{CIRCL}}}
\newcommand{\awslc}{\textbf{\texttt{AWS-LC}}}
\newcommand{\pqm}{\textbf{\texttt{pqm4}}}
\newcommand{\libtomcrypt}{\textbf{\texttt{libtomcrypt}}}
\newcommand{\pysnark}{\textbf{\texttt{pysnark}}}
\newcommand{\lattice}{\textbf{\texttt{lattice\_zksnark}}}

\newcommand{\pqclean}{\textbf{\texttt{PQClean}}}
\newcommand{\libpq}{\textbf{\texttt{libpqcrypto}}}

\newlist{Qenumerate}{enumerate}{1}
\setlist[Qenumerate]{label={Q}\arabic*}
\newlist{Cenumerate}{enumerate}{1}
\setlist[Cenumerate]{label={C}\arabic*}

\begin{document}

\title{Classifying Implementations of Cryptographic Primitives and Protocols that Use Post-Quantum Algorithms}

\author{
\IEEEauthorblockN{Tushin Mallick}
\IEEEauthorblockA{
Northeastern University\\
mallick.tu@northeastern.edu
}
\and
\IEEEauthorblockN{Cristina Nita-Rotaru}
\IEEEauthorblockA{
Northeastern University\\
c.nitarotaru@northeastern.edu
}
\and
\IEEEauthorblockN{Ashish Kundu}
\IEEEauthorblockA{
Cisco\\
ashkundu@cisco.com
}
\and
\IEEEauthorblockN{Ramana Kompella}
\IEEEauthorblockA{
Cisco\\
rkompell@cisco.com
}
}

\maketitle

\raggedbottom

\begin{abstract}
% Fingerprinting is a technique used to create 
% profiles of systems to identify potential threats, weaknesses, and  malicious activities. For cryptographic primitives and  network protocols such fingerprinting can be exploited by attackers to conduct denial-of-service, key recovery, or downgrade attacks.

Classification techniques can be used to analyze system behaviors, network protocols, and cryptographic primitives based on identifiable traits. While useful for defensive purposes, such classification can also be leveraged by attackers to infer system configurations, detect vulnerabilities, and tailor specific attacks — including denial-of-service, key recovery, or downgrade attacks — against the categorized targets.

In this paper, we study the feasibility of classifying post-quantum (PQ) algorithms by analyzing implementations of primitives such as
key exchange and digital signatures, their usage within secure protocols, and integration into SNARK generation libraries.
Unlike traditional cryptographic algorithms, PQ algorithms have larger memory needs and variable computation costs. Our research examines two post-quantum cryptography libraries, \liboqs~ and \circl. We evaluate their implementations of TLS, SSH, QUIC, OpenVPN, and OpenID Connect (OIDC) across Windows, Ubuntu, and macOS. Additionally, we analyze two SNARK generation and verification libraries—\pysnark~ and \lattice—on Ubuntu. Experimental results reveal that, depending on computational workload, (1) we can distinguish between classical and PQ key exchange and digital signature algorithms with accuracies of 98\% and 100\%, respectively; (2) we can determine the used algorithm within PQ key exchange or digital signature algorithms with 97\% and 86\% accuracy, respectively; (3) we can distinguish between implementations of the same algorithm in the \liboqs\ and \circl\ libraries for key exchange and digital signature with 97\% and 100\% accuracy, respectively; and (4) within the same library --\circl\, we can distinguish the PQ and hybrid key exchange algorithm implementations with 97\% accuracy. In the case of secure protocols, it is possible to (1) discern whether the key exchange is classical or PQ, and (2) identify the specific PQ key exchange algorithm employed. Finally, the SNARK generation and verification of \pysnark\ and \lattice~ are distinguishable with an accuracy of 100\%. We demonstrate the applicability of our classification methods to real systems by applying them to the Tranco dataset to identify domains that use PQ key exchange for TLS, and by integrating them in QUARTZ, an open source risk and threat analyzer created by CISCO.
\end{abstract}

\section{Introduction}
\label{sec:introduction}

Quantum computing leverages quantum mechanics to process data in fundamentally new ways, enabling exponentially faster solutions to complex problems than classical computers. A major area of impact is cryptography, as schemes like RSA and Diffie-Hellman rely on computational problems vulnerable to quantum attacks. Although practical quantum computers are not yet available, researchers are developing cryptographic methods resilient to future quantum threats and designed to prevent ``harvest-now, decrypt-later'' attacks.

NIST conducted a multi-year initiative to evaluate and standardize cryptographic algorithms resistant to quantum attacks, leading to the establishment of post-quantum cryptography (PQC)~\cite{nist-pqc-details}. The process concluded in 2023 with public drafts for encryption~\cite{nist-fips203_encryption} and digital signatures~\cite{nist-fips204_ds,nist-fips205_hds}, followed by the release of the first PQC standards—FIPS 203~\cite{NIST-pqc-standard-fips203}, FIPS 204~\cite{NIST-pqc-standard-fips204}, and FIPS 205~\cite{NIST-pqc-standard-fips205}—in August 2024~\cite{NIST-pqc-standard-Aug132024}. In March 2025, NIST selected Hamming Quasi-Cyclic (HQC) as a fifth key encapsulation mechanism, providing a non-lattice alternative to ML-KEM (FIPS 203), with a draft standard expected in 2026 and finalization by 2027. Concurrently, draft NIST IR 8547, Transition to Post-Quantum Cryptography Standards, offers guidance on migration planning, risk assessment, crypto agility, and asset inventory, recommending deprecation of quantum-vulnerable algorithms (RSA, ECDSA, EdDSA, DH, ECDH) by 2030 and their full retirement by 2035.

Implementations of post-quantum cryptographic (PQC) primitives are now publicly available in libraries and integrated into secure communication protocols and storage systems. Notable examples include \liboqs~\cite{liboqs}, \circl~\cite{circl}, \awslc~\cite{aws-lc}, \pqm~\cite{pqm4}, \pqclean~\cite{pqclean}, and \libpq~\cite{libpqcrypto}. \liboqs, developed by the Open Quantum Safe project under the Linux Foundation’s Post-Quantum Cryptography Alliance, supports integration and evaluation of PQC algorithms across major operating systems. \circl, by Cloudflare, provides advanced cryptographic algorithms for the Go programming language. Amazon’s \awslc~library incorporates PQC schemes to strengthen AWS security. The \pqm~library targets resource-constrained embedded systems on ARM Cortex-M4. \pqclean~offers clean, portable C implementations of NIST PQC candidates for reference and reuse, while \libpq, from the PQCRYPTO project, collects diverse encryption and signature schemes for research and benchmarking.

Additionally, it has become crucial to secure protocols like QUIC, SSH, DTLS \cite{rfc9147}, TLS, and OAuth 2.0 against quantum threats. Major tech companies are already adopting hybrid and quantum-safe solutions: Google, Microsoft, Cloudflare, and Amazon integrate quantum algorithms into their protocols; IBM offers quantum-safe TLS; Meta uses hybrid key exchange; Akamai prepares PQC deployment across its CDN; and Cisco embeds quantum-safe methods in its networking products—signaling an industry-wide shift toward quantum-resilient infrastructure.

Post-quantum algorithms have distinct memory footprints and CPU cycle patterns, due to their complex operations, larger keys, and ciphertexts. These characteristics make them identifiable through classification techniques that analyze behavioral patterns and protocol-specific responses. Such classification can expose systems to risks, as attackers may exploit identifiable traits to launch denial-of-service, key recovery, or downgrade attacks. Conversely, it can aid defenders in assessing risks and identifying vulnerable components. Previous research has focused on classifying secure protocols using classical cryptography, underscoring the growing importance of similar approaches for post-quantum systems.

% Xue et al. \cite{xue2022} demonstrated how the network flow of OpenVPN could be fingerprinted by identifying fixed patterns within OpenVPN traffic. Similarly, Panchenko et al. \cite{torf} presented an attack on Tor, revealing that traffic analysis could be used to identify visited websites, despite the use of encryption, by exploiting unique traffic patterns generated by different websites. Wright et al. and Wang et al\cite{wright2006inferring, 1dcnn} have shown that machine learning models applied to encrypted traffic can effectively identify the underlying protocols, highlighting vulnerabilities in encrypted communications.

\textbf{Our contribution. }In this work we investigate whether post-quantum cryptographic primitives can be classified based on their library implementations, use in secure communication protocols, or integration in advanced constructions like SNARKs. The classification task is challenging because post-quantum algorithms exhibit diverse computational and memory characteristics—some requiring more resources, others less, and some closely matching classical algorithms. These inconsistencies make rule-based approaches unreliable, motivating the use of machine learning models to capture subtle behavioral differences between classical and post-quantum schemes.

% In this work we explore if it is possible to classify post quantum primitives from their library implementations, their usage within protocols that protect Internet communication, or more complex cryptographic constructions such as SNARK-s widely used in blockchains. The classification of post-quantum and classical algorithms is inherently non-trivial, as computational time and memory usage do not provide consistent discriminative features. In some instances, post-quantum algorithms may require either greater or fewer resources than classical algorithms, and certain variants even exhibit comparable resource consumption to their classical counterparts. Owing to this complexity, rule-based filtering is insufficient for reliable classification, thereby necessitating the use of machine learning models to capture the nuanced distinctions between the two categories. 

We start with publicly available post quantum libraries from Microsoft, Cloudflare, Amazon to show that, when compared to classical algorithms in terms of computational cost and memory overhead their post quantum counterparts are easily distinguishable by using machine learning models. We collected data from \liboqs\ and \circl\ running on Windows, Ubuntu, and MacOS under a scenario when other intensive computation tasks share the resources with the post quantum libraries. We formulate distinguishing key exchange schemes as well as signature schemes, employing machine learning methods to carry out the classification. Ensemble learning models, particularly XGBoost, exhibit the highest performance across various classification tasks. Specifically, it achieved 98\% accuracy in classifying classical vs. post quantum key exchange algorithms and 100\% accuracy for classical vs. post-quantum digital signatures. For post-quantum key exchange algorithms, the classification accuracy reached 97\%, while for post-quantum digital signatures, it was 86\%. XGBoost also effectively distinguished between implementations of the same algorithm in the \liboqs\ and \circl\ libraries, achieving 96\% accuracy in key exchange and 100\% in digital signature. For post-quantum and hybrid key exchange classification within \circl\ library, it achieved 97\% accuracy.

%Once we reach the conclusion that post quantum and classical algorithms are computationally very distinct, 
We next proceed to see if classification is possible when post quantum algorithms are implemented for key exchanges in widely used protocols such as TLS\cite{tls}, SSH\cite{rfc4253}, QUIC\cite{rfc9000}, OpenVPN \cite{openvpn} and Open ID Connect(OIDC) \cite{rfc6749, rfc6750}. We start by gathering packet captures for these protocols with Wireshark and tcpdump. For classical packet captures we either use publicly available captures from Cloudflare\cite{qacafe_sample_captures}, Wireshark \cite{wireshark_sample_captures} or use tcpdump on existing classical implementations of those protocols to generate packet captures locally in our devices. For protocols using TLS, such as OIDC, QUIC, and OpenVPN, we isolate the key exchange packets and then compare the key sizes within the TLS layer to determine whether the connection employs a classical or post quantum algorithm. Similarly, for SSH, we isolate the packets and identify key exchange from information embedded in the SSH layer instead of TLS. By integrating these approaches, we successfully identify the key exchange algorithms in the majority of TLS 1.3 and all SSH connections with the exception of  a small number of connections such as certain Windows TLS connections using custom headers and fragmented TLS 1.2 connections from OpenVPN.

Finally, we study the feasibility of classifying SNARK generation and verification in the \pysnark~ (classical) and \lattice~ (post-quantum) libraries.
%, due to their use in numerous blockchain applications. %The findings indicate that, in terms of computational cost and memory overhead, post-quantum libraries are readily distinguishable from their classical counterparts based on computation time. 
XGBoost is the most effective, achieving 100\% accuracy between the post-quantum and classical library implementations.

We demonstrate the applicability
of our classification methods to real systems by applying
them to the Tranco dataset to identify domains that use PQ
key exchange for TLS, and by integrating them in QUARTZ,
an open source risk and threat analyzer created by CISCO.

\begin{comment}
In summary, our key contributions are as follows:
\begin{itemize}
 %   \item We provide an overview of existing implementations of post-quantum building blocks and secure communication protocols using them.
    \item We demonstrate the feasibility of identifying classical and post quantum key exchange, digital signature schemes, as well as the libraries in which they are implemented.
    \item We demonstrate the ability to identify both the key exchange mechanism—post-quantum or classical—and the specific algorithm used by a connection in well-known and widely-used secure protocols.
    \item We demonstrate the feasibility of binary classification to distinguish between classical and post-quantum integration within SNARK generation libraries. 
    \item We demonstrate the applicability of our techniques to real systems by integrating them in QUARTZ, an open source enterprise quantum risk and threat analyzer created by CISCO.
    \item We isolate domains from the Tranco dataset which use post quantum key exchange for TLS handshakes
\end{itemize}
\end{comment}

\textbf{Ethics.} We collected the data on our own personal laptops. We do not plan to make the data available, but we provide scripts for data collection and code to reproduce our results.  

\textbf{Availability.}
%-------------------------------------------------------------------------------
Our code is available at \url {https://anonymous.4open.science/r/Fingerprint-809D/README.md}.

\section{Background}
\label{sec:pq}

Post-quantum (PQ) cryptography is crucial to ensuring security in the face of emerging quantum computing threats. As quantum computers advance, they pose a risk to classical cryptographic algorithms, making the adoption of PQ key exchange mechanisms and digital signature schemes essential. Gradually, these PQ algorithms are being integrated into existing protocols to enhance security and future-proof systems against quantum attacks. Below
we overview current PQ algorithms and some protocols that implement them.

\subsection{KEM and Digital Signature Algorithms}

The algorithms are categorized based on the mathematical problems they rely on, each offering different strengths, weaknesses, and applications. A summary of the algorithms is provided in Table \ref{tab:postquantum} and we expound on the categories below. 

\textbf{Lattice-based} cryptography is one of the most promising and versatile areas of post-quantum cryptography. It relies on problems in lattice theory, such as the Shortest Vector Problem (SVP) and the Learning With Errors (LWE) problem. These problems are believed to be difficult for both classical and quantum computers to solve. Lattice-based cryptographic schemes are valued for their efficiency in computation and storage, making them suitable for a broad range of applications, including key exchange, encryption, and digital signatures. Key exchange algorithms CRYSTALS-Kyber\cite{kyber2018}, NTRU\cite{ntru1998}, SABER\cite{saber2018} and digital signature scheme CRYSTALS-Dilithium are finalists in the NIST post-quantum cryptography standardization process.

\textbf{Code-based} cryptography is rooted in the hardness of decoding random linear codes—a problem considered difficult even for quantum computers.  While this category of algorithms offer small ciphertexts and fast encryption/decryption, their main drawback is the large size of public keys, which can be challenging to manage. Key exchange algorithm Classic McEliece\cite{mceliece1978} is a finalist in the NIST post-quantum cryptography standardization process for this category.

\textbf{Hash-based} cryptography focuses on the security provided by cryptographic hash functions. These schemes are particularly attractive for digital signatures because their security  does not  rely on complex mathematical structures like lattices but instead is directly tied to the difficulty of finding hash collisions, a problem quantum computers do not solve much faster than classical computers. Hash-based cryptography is highly secure but typically involves larger signature sizes.

\textbf{Multivariate} polynomial cryptography is based on the difficulty of solving systems of multivariate quadratic equations over finite fields, a problem known to be NP-hard. Algorithms in this category, such as Rainbow\cite{rainbow2005} and GeMSS\cite{gemss2018}, are primarily used for digital signatures. These schemes provide a valuable alternative for post-quantum security, despite their complexity and inefficiency due to large keys and signatures. Rainbow is a finalist in the NIST post-quantum cryptography standardization process for this category.

\textbf{Isogeny-based} cryptography represents a more specialized area, using the mathematical structure of elliptic curves and the difficulty of computing isogenies between them. Schemes like SIKE (Supersingular Isogeny Key Encapsulation) \cite{sike2017} are compact and have relatively small key sizes, which is a significant advantage for certain applications like secure communications in constrained environments. However, in July 2022, KU Leuven researchers broke the SIKE algorithm, a NIST fourth-round candidate, using a classical computer in 62 minutes, highlighting vulnerabilities in its underlying supersingular isogeny problem.

\textbf{Zero-knowledge} proof-based post-quantum algorithms are designed to enable secure verification processes without revealing the underlying data, even in the face of quantum computing threats. 

\textbf{Hybrid Algorithms.} Post-quantum hybrid algorithms combine traditional cryptographic methods with post-quantum cryptography to provide enhanced security during the transition to a quantum-resistant future. Technically, these algorithms generate two distinct key pairs: one from a well-established classical cryptosystem, such as RSA or Elliptic Curve Cryptography (ECC), and another from a post-quantum cryptosystem, like those mentioned above. The hybridization ensures that even if quantum computers become powerful enough to break classical encryption, the post-quantum components will still safeguard the data. As the cryptographic community continues to develop and standardize these algorithms, hybrid solutions will play a vital role in protecting sensitive information against current and future threats.

\textbf{Standardization.} 
Earlier in 2022, NIST selected four algorithms: CRYSTALS-Kyber, CRYSTALS-Dilithium, Sphincs + and FALCON for standardization, with draft versions of three standards released in 2023. In August 2024, NIST released the first PQC standards ~\cite{NIST-pqc-standard-Aug132024}: FIPS 203~\cite{NIST-pqc-standard-fips203}, FIPS 204~\cite{NIST-pqc-standard-fips204}, and FIPS 205~\cite{NIST-pqc-standard-fips205}. NIST is working to standardize a second set of algorithms.
The fourth draft standard, based on FALCON, was expected in late 2024. But as of 2025, there has been no announcement of a final release yet.
%Although there have been no substantive changes since the draft releases, 

NIST has renamed the algorithms to reflect their finalized versions in three Federal Information Processing Standards (FIPS). FIPS 203, based on CRYSTALS-Kyber, is now named ML-KEM (Module-Lattice-Based Key-Encapsulation Mechanism) and serves as the primary standard for general encryption. FIPS 204, derived from CRYSTALS-Dilithium, is renamed ML-DSA (Module-Lattice-Based Digital Signature Algorithm) and is the primary standard for digital signatures. FIPS 205, using Sphincs+, is renamed SLH-DSA (Stateless Hash-Based Digital Signature Algorithm) and serves as a backup for digital signatures, employing a different mathematical approach. The forthcoming FIPS 206, based on FALCON, will be named FN-DSA (FFT over NTRU-Lattice-Based Digital Signature Algorithm).

\begin{table*}[h!]
\centering
\scriptsize
\setlength{\tabcolsep}{4pt} % Adjust space between columns
\renewcommand{\arraystretch}{1.2} % Adjust space between rows
\begin{tabular}{|p{3cm}|p{3cm}|p{4cm}|p{3cm}|p{3cm}|}
\hline
\textbf{Category} & \textbf{Name of Algorithm} & \textbf{Variants} & \textbf{Type} & \textbf{Implementations} \\ \hline

\multirow{2}{*}{Lattice-based} 
& CRYSTALS-Kyber & Kyber512, Kyber768, Kyber1024 & Key Encapsulation & liboqs\cite{liboqs}, CIRCL\cite{circl}, pqm4\cite{pqm4}, AWS-LC \cite{aws-lc} \\ \cline{2-5}
& FRODOKem\cite{frodo2016} & FRODO-640, FRODO-976, FRODO-1344 & Key Encapsulation & liboqs, CIRCL \\ \cline{2-5}
&SABER &  LightSABER, SABER, FireSABER & Key Encapsulation & N/A \\ \cline{2-5}
& NTRU & NTRUEncrypt, NTRU-HRSS-KEM, and NTRU Prime & Key Encapsulation & liboqs, CIRCL \\ \cline{2-5}
& CRYSTALS-Dilithium & Dilithium-2, Dilithium-3, Dilithium-5 & Digital Signature & liboqs, CIRCL, pqm4 \\ \cline{2-5}
& FALCON \cite{FALCON2018} & FALCON-512, FALCON-1024 & Digital Signature & liboqs, pqm4\\ \cline{2-5}

& ML-DSA\cite{ml-dsa} & ML-DSA-44, ML-DSA-65, ML-DSA-87 & Digital Signature & liboqs\\ \hline

\multirow{2}{*}{Code-based} 
& Classic McEliece & Classic-McEliece-348864, Classic-McEliece-460896, Classic-McEliece-6688128, Classic-McEliece-6960119, Classic-McEliece-8192128 & Key Encapsulation & liboqs\\ \cline{2-5}
& BIKE \cite{bike2021} & BIKE-L1, BIKE-L3, BIKE-L5 & Key Encapsulation & liboqs, pqm4 \\ \cline{2-5}
& HQC\cite{hqc2018}  & HQC-128, HQC-192, HQC-256 & Key Encapsulation & liboqs \\ \hline

\multirow{2}{*}{Hash-based} 
& SPHINCS+ \cite{sphincs2019} & SPHINCS+-SHA2-128-simple,
SPHINCS+-SHA2-192-simple, SPHINCS+-
SHA2-256-simple & Digital Signature & liboqs \\ \cline{2-5}
&XMSS\cite{xmss2011} & XMSS, XMSS-MT & Digital Signature & N/A \\ \hline

\multirow{2}{*}{Multivariate polynomial based} 
& Rainbow & Rainbow-I, Rainbow-III, Rainbow-V & Digital Signature & N/A \\ \cline{2-5}
& GeMSS & GeMSS128, GeMSS192, GeMSS256 & Digital Signature & N/A\\ \hline

\multirow{1}{*}{Isogeny-based} 
& SIKE & SIKE-p434, SIKE-p503, SIKE-p610 & Key Encapsulation & liboqs\\ \hline

\multirow{2}{*}{Zero Knowledge-based}
& PICNIC\cite{picnic2019} & PICNIC2-L1-FS, PICNIC2-L3-FS, PICNIC2-L5-FS & Digital Signature & N/A \\ \hline

\end{tabular}
\caption{Post-Quantum Cryptography Algorithms}
\label{tab:postquantum}
\end{table*}

\subsection{SNARKs}

Succinct Non-Interactive Arguments of Knowledge or SNARKs are cryptographic constructs enabling a prover to convince a verifier of computation correctness without revealing computation details. 
Their non-interactive nature enhances scalability, privacy, and efficiency, particularly in blockchains for privacy-preserving transactions and computational efficiency.
SNARKs also have applications in secure cloud computing and confidential ML.

SNARKs involve three components: a generator that creates a Common Reference String (CRS) for shared parameters, a prover that encodes the computation as constraints and generates a proof, and a verifier that validates the proof without evaluating the computation directly. SNARK security is evaluated quantitatively by the computational effort required to forge a false proof, measured in ``bits of security''. For instance, a SNARK with 40 bits of security means an attacker would need $2^{40}$ operations to forge a proof. Qualitative security is the probability of information leakage, ensuring zero-knowledge under statistical assumptions. For example, a qualitative security level of 20 means that the chance of an adversary successfully breaking the zero-knowledge guarantee (e.g., deducing some information about the private input) is 
$2^{-20}$. Many SNARKs (e.g., Groth16, PlonK, Marlin, Bulletproofs, Nova) rely on discrete logarithm hardness, which quantum algorithms can efficiently solve, rendering them non-post-quantum secure.

Libraries such as lattice-zksnark\cite{ISW21} are actively working toward implementing PQ SNARKs by utilizing parameters resistant to quantum attacks. For example, lattice-zksnark uses a quantitative security of 128 bits and a qualitative security of 40. Other efforts in this domain include zkLLVM\cite{zkllvm}, which explores compiling SNARK-friendly circuits with post-quantum backends, and research prototypes based on STARKs (Scalable Transparent Arguments of Knowledge), which use hash-based security and do not rely on discrete logarithms. Solana, a high performance blockchain introduced the ``Solana Winternitz Vault'', a quantum-resistant, hash-based signature system to secure user funds against quantum threats. This optional feature regenerates private keys per transaction, enhancing security for those who opt in.

\subsection{Integration of Post Quantum KEM and Digital Signature in Secure Protocols}

%The integration of post-quantum Key Encapsulation Mechanisms (KEMs) and digital signatures is essential for ensuring the long-term security of digital communications against quantum threats. As quantum computers threaten traditional cryptography like RSA and ECC\cite{ecc1985, koblitz1987}, transitioning to post-quantum cryptographic (PQC) solutions is critical to safeguarding sensitive data and securing protocols. 
%Efforts are underway to adopt post-quantum algorithms, ensuring resilience against emerging quantum computing capabilities.

The adoption of PQ algorithms in widely used secure protocols, such as TLS, QUIC, VPN, Open Id Connect (OIDC), and SSH, is critical to mitigating future quantum threats. These protocols are transitioning through hybrid approaches, combining classical and PQ cryptography for backward compatibility. For instance, TLS and QUIC are integrating PQ key exchanges, VPNs are employing PQ KEMs for encrypted tunnels, OIDC is updating identity token signing and TLS handshakes with PQ signatures and key exchanges, and SSH is incorporating PQ algorithms in its key exchange processes. This phased approach allows existing systems to maintain security as they gradually shift to post-quantum cryptographic standards. We provide more details in Table \ref{tab:networking_security}.

\textbf{Hybrid Key Exchange in TLS 1.3.} Every TLS connection starts with a handshake which is necessary to securely establish a shared encryption key and negotiate encryption protocols before any data is exchanged, ensuring the confidentiality and integrity of the connection. It begins with the ClientHello message, where the client proposes cryptographic settings for key exchange. TLS 1.3 ClientHello messages always include an extensions field (minimally ``supported\_versions'') which contains the public keys of the algorithms the client supports. This is followed by the ServerHello message, where the server agrees on the parameters. The server then sends a Server Certificate and may request a client certificate. Finally, the client and server exchange Finished messages after generating session keys, completing the handshake and allowing secure data transmission.
%\textbf{Example: TLS post-quantum and hybrid key exchange}
Below we describe how TLS key exchange is changed in the PQ and hybrid scenarios.

\begin{figure}[h!]
  \centering
  \includegraphics[width=\linewidth, trim=0cm 1cm 1cm 0cm]{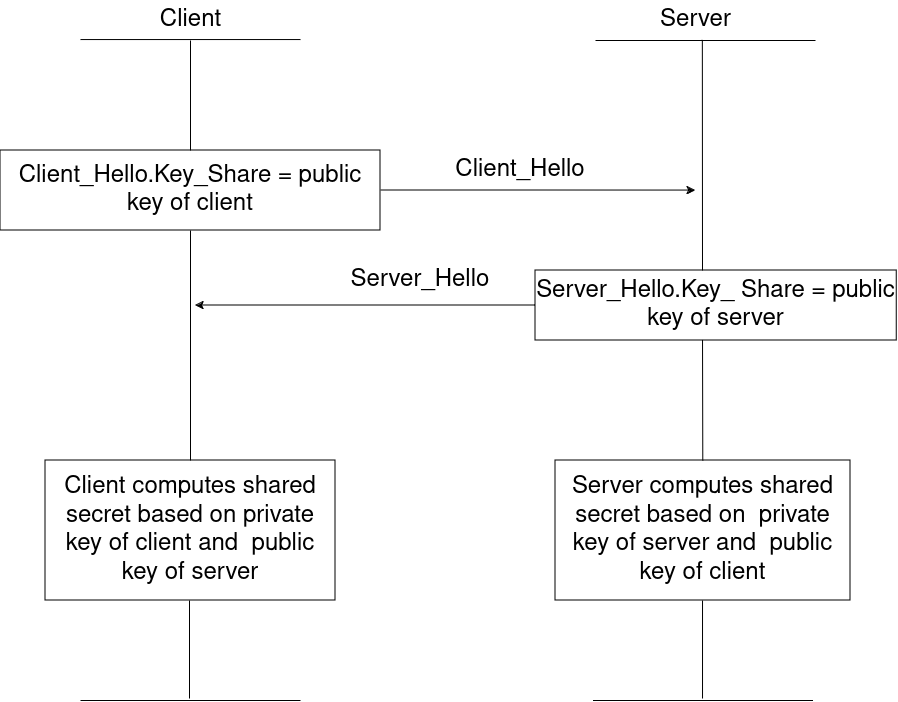} 
  \caption{Simplified Classical Key Exchange in TLS 1.3}
  \label{fig:classical key exchange}
\end{figure}

\textit{Classical}. In TLS 1.3, the client initiates the  Diffie-Hellman key exchange by selecting a secret key \textit{x} and computing the corresponding public key $g^{x}$, which is then included in the extension field of the ClientHello. Upon receiving this, the server selects its own secret key \textit{y} and computes its public key $g^{y}$, transmitting it in the ServerHello. The shared secret $g^{xy}$ is derived independently by both the client and server, enabling secure communication (see Figure \ref{fig:classical key exchange}).

\textit{Post-quantum.} 
In post quantum key exchange the public key of the post quantum algorithm is transmitted as an entry of the ClientHello extensions field. The server then generates a random secret, encrypts it with the client’s post-quantum public key, and sends the ciphertext via the ServerHello. The client decrypts the ciphertext using its post-quantum private key to retrieve the post-quantum shared secret.

\begin{figure}[h!]
  \centering
  \includegraphics[width=\linewidth, trim=0cm 1cm 1cm 2cm]{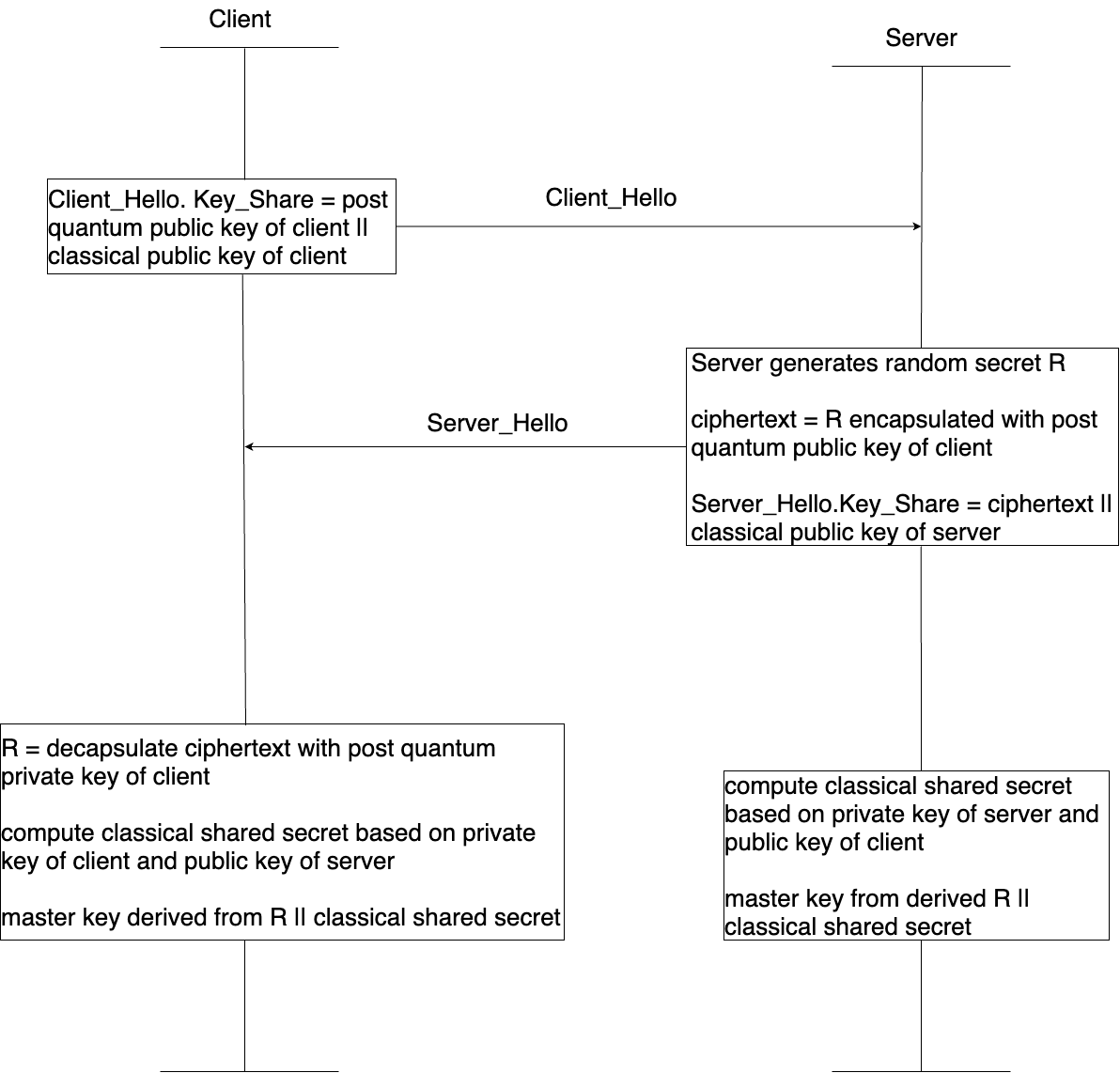} 
  \caption{Hybrid Key Exchange in TLS 1.3}
  \label{fig:hybrid key exchange}
\end{figure}

\textit{Hybrid.}
In a hybrid key exchange within TLS 1.3, both classical and post-quantum keys are generated. The classical key is derived from methods like Diffie-Hellman (DH) or Elliptic Curve Diffie-Hellman (ECDHE), while the post-quantum key is obtained from a quantum-resistant algorithm. The client’s ClientHello message contains the concatenated post-quantum and classical public keys as an entry of the extensions field. These keys are concatenated without any additional encoding and transmitted as a single value, avoiding alterations to existing protocol data structures. The server then generates a random secret, encrypts it with the client’s post-quantum public key, and sends the ciphertext along with its classical public key in the ServerHello. The client decrypts the ciphertext using its post-quantum private key to retrieve the post-quantum shared secret, while the classical shared secret is computed using the classical key exchange method. The final shared secret is derived by concatenating the results of both the classical and post-quantum exchanges (see Figure \ref{fig:hybrid key exchange}). This information is sourced from the draft available at \cite{draft-ietf-tls-hybrid-design-10}.

\begin{table*}[h!]
\centering
\scriptsize
\setlength{\tabcolsep}{10pt} % Adjust space between columns
\renewcommand{\arraystretch}{1.2} % Adjust space between rows
\begin{tabular}{|p{2.3cm}|p{7cm}|p{2cm}|p{2cm}|p{0.9cm}|}
\hline
\textbf{Name} & \textbf{Description} & \textbf{Classic Implementation} & \textbf{PQ Implementation}  & \textbf{PQ libraries used}\\ \hline

\textbf{TLS (Transport Layer Security)} & TLS operates on top of the TCP layer to secure data over networks with encrypted handshakes to authenticate parties and establish secure channels. \textbf{Post-quantum} handshakes replace classical methods with post-quantum or hybrid keys. &  OpenSSL\cite{openssl} \newline Schannel \cite{schannel}\newline Secure Transport API\cite{secure-transport} & s2n-TLS\cite{s2n-tls}\newline OQS-OpenSSL\cite{oqs-openssl}\newline wolfSSL\cite{wolfssl_pqc}\newline & AWS-LC\cite{aws-lc}\newline liboqs\newline wolfCrypt 
 \\ \hline

\textbf{QUIC (Quick UDP Internet Connections)} & QUIC, a transport protocol by Google, enhances communication with 0-RTT connections, TLS 1.3 encryption, multiplexing, and seamless migration. \textbf{Post-quantum} handshakes use post-quantum or hybrid keys instead of classical ones.  & Quiche\cite{quiche}\newline gQUIC\cite{gquic} & s2n-Quic\cite{s2n-quic} & AWS-LC
\\ \hline

\textbf{VPN (Virtual Private Network)} & VPNs create secure, encrypted tunnels over the internet, protecting data and ensuring privacy during remote network access or internet use. They use protocols like IPsec or TLS, with TLS securing initial connections. \textbf{Post-quantum} handshakes replace classical keys with post-quantum or hybrid keys to counter quantum threats. & OpenVPN\cite{openvpn} Wireguard\cite{wireguard} L2TP/IPSEC\cite{l2tp-ipsec} IPSEC\cite{ipsec}\newline PPTP\cite{pptp}\newline SSTP\cite{sstp} & PQ-crypto-OpenVPN\cite{pq-openvpn}\newline PQ-Wireguard\cite{pq-wireguard} & liboqs
\\ \hline

\textbf{OIDC (OpenID Connect)} & OpenID Connect (OIDC), built on OAuth 2.0, enables clients to verify user identities via an authorization server. The OIDC Provider (OP) issues ID tokens, which Relying Parties (RP) use to authenticate users and grant access. \textbf{Post-quantum} TLS handshakes secure communication, while the OP employs post-quantum signatures for signing JWT tokens, protecting against quantum threats. & Auth0\cite{auth0}\newline Okta\cite{okta}\newline Google Identity\cite{google-identity}\newline Entra ID\cite{entra-id} & Post-quantum-oidc-oauth2 \cite{postquantum-oidc-oauth2} & liboqs
\\ \hline

\textbf{SSH (Secure Shell)} & SSH is a protocol for securely accessing and managing remote servers via encrypted communication. It enables secure command execution, file transfers, and port forwarding. \textbf{Post-quantum} handshakes replace classical key exchanges with post-quantum or hybrid keys to counter quantum threats. & OpenSSH (version < 9.0)\cite{openssh}\newline PuTTY\newline BitVise\newline Tectia SSH & OpenSSH (version >= 9.0)\newline PQ-OpenSSH  & liboqs\\ \hline

\end{tabular}
\caption{Secure Protocols Supporting Post-Quantum Algorithms}
\label{tab:networking_security}
\end{table*}

\section{Classification of Libraries}

In this section, we describe the classification of key exchange algorithms and signature schemes in library implementations. We first describe our methodology, then we describe our algorithm, then describe the detection for key exchange and digital signatures. 
We seek to answer the following questions: 
\begin{itemize} 

\item \textit{RQ1: Is it possible to distinguish between classical and post quantum algorithms?}

\item \textit{RQ2: Within a specific library, is it possible to distinguish between different post-quantum algorithms?}

\item \textit{RQ3: Is it possible to distinguish post quantum algorithm implementations across different libraries?}

\item \textit{RQ4: Is it possible to distinguish between basic post quantum and hybrid key exchange algorithms?}

\end{itemize}

\subsection{Methodology}

\textbf{Libraries we consider.} We run experiments on publicly available libraries such as \liboqs~ and \circl~ that implement PQ key exchange and signature algorithms. We did not utilize \pqm~ in our experiments because it is specifically optimized for the ARM Cortex-M4 microcontroller, which is primarily used in embedded systems rather than general-purpose computing environments. Our experiments focused on evaluating PQ cryptographic algorithms within operating systems such as Ubuntu, macOS, and Windows, which are representative of desktop and server environments where the broader deployment of cryptographic protocols is more relevant. Integrating \pqm~ alongside the other libraries would yield results that are not directly comparable or representative of the broader deployment contexts of cryptographic protocols. We did not consider \awslc\ library because it has only 1 implementation of CRYSTALS-Kyber. The \pqclean~ project, although providing verified and portable reference implementations, covers only a limited subset of PQ algorithms relative to \liboqs~ and \circl. Incorporating it would have introduced sampling bias toward that restricted subset. We also attempted to compile \libpq, but the build process consistently failed due to compile-time errors, preventing its inclusion in the benchmark suite.
% Finally, \wolf~ was excluded because its PQ algorithms are embedded within higher-level protocol implementations such as TLS, rather than being exposed as standalone cryptographic primitives. 
Implementations of hybrid algorithms were only available in \circl. Classical implementations of algorithms were taken from the \libtomcrypt~ library. The \liboqs~ and \libtomcrypt~ libraries are written in C, while \circl~ is written in Go. 

\textbf{Available algorithms.} Classical key exchange algorithms include RSA\cite{rsa1978} and ECDH variants. Classical signature schemes include variants of DSA\cite{rfc6979}, RSA and ECDSA\cite{ecdsa}.

PQ key exchange algorithms include variants of CRYSTALS-Kyber, BIKE, FRODO, HQC, Classic McEliece, NTRUPrime from \liboqs~ and variants of FRODO, CRYSTALS-Kyber, SIKE from \circl. PQ digital signature schemes include variants of CRYSTALS-Dilithium, FALCON, ML-DSA, SPHINCS+ from \liboqs~ and variants of CRYSTALS-Dilithium from \circl.

Hybrid key exchange algorithms include Kyber512 + X25519, Kyber768 + P256, Kyber768 + X448, Kyber768 + X25519, Kyber1024 + X448 from \circl~ and \liboqs~ lacks implementations for any hybrid algorithm. We are not aware of any hybrid digital signature algorithms.
 %Hybrid algorithms from all libraries are excluded from our experiments because, if fingerprinting is feasible for a given post-quantum algorithm, it will also be feasible within a hybrid scheme that incorporates the algorithm, as hybrid schemes maintain the operational integrity of both their classical and post-quantum components. The primary difference lies in the additional computational and memory overhead, which will in turn ease our fingerprinting.

\textbf{Device specifications.} Experiments were conducted on three different computing devices. The first device was equipped with Ubuntu 22.04, powered by an Intel Core i5-11400H CPU, which features 6 physical cores and 12 virtual cores, complemented by 16 GB of RAM. The second device operated on Windows 10, utilizing an Intel Core i5-6300U CPU with 2 physical cores and 4 virtual cores, also paired with 16 GB of RAM. The third device, also running Windows 10, was configured with an Intel Core i5-4300U CPU, with 2 physical cores and 4 virtual cores, and 8 GB of RAM.

\textbf{Datasets.} CPU cycle count and memory usage were considered as features for the classification tasks. The libraries have dependencies on system calls unique to Linux systems and so cannot be carried out solely on Windows without WSL\cite{wsl}. The \textbf{\texttt{perf}} command was used to get a per core cycle count of the running process. Although \textbf{\texttt{perf}} inherently reports system-wide cycle counts, background activity was minimized to ensure that the reported measurements provided a sufficiently accurate approximation of the per-core cycle count attributable to the running process. For memory usage we captured the memory imprint of a running process from the /proc/[pid]/status file in Linux. The cycle counts and memory usage for the processes were generated across 1000 runs. 

To mimic real world scenarios we segregate our data for each classification task into two types of datasets. \textbf{\datasetA} contains data where we minimize running background processes as much as possible and \textbf{\datasetB} where we performed in the background tasks with CPU intensive operations such as matrix multiplications, prime number calculations, sorting of large lists and hash calculations run in parallel. Running additional tasks impacted the computation time, but did not affect the memory usage of the tested process.
%as the status file in Linux is unique to the pid of the process.

\subsection{Classification Method} 
We formulate the problem as a machine learning classification, with classes identified by the names of the cryptographic algorithms. For each classification scenario in key exchange, we start the classification task with 19 features of which 12 features represent the cycles used by each core of the CPU and remaining 7 features represent a process's memory usage, including total virtual memory(VMSize), physical memory in use(VMRSS), data segment(VMData), stack, executable code(VMStk), shared libraries(VMLib), and page table entries(VMPTE). For devices having less than 12 virtual CPU cores the additional fields were kept empty and we did not have devices with more than 12 cores. We narrow down our feature set to 4 memory based features (VMSize, VMRSS, VMData, VMExe) using the Chi-square statistical test as the scoring function to get the top features by importance. In digital signatures, we used the same 4 features in our set for two classification scenarios but for the multi-class scenario we needed all 19 features. We do an 80-20 train-test split of the dataset and fit models such as Logistic Regression, Random Forest, MLP and XGBoost to the data.

\subsection{Key Exchange Results}

This section presents our findings across four classification scenarios, which form the basis for answering to questions related to the classification of PQ  key exchange algorithm implementations in libraries. We begin with the classification between classical and PQ algorithms, followed by a multi-class classification focused exclusively on PQ algorithms. Next, we distinguish between the two libraries used in our experiments based on their algorithm implementations. Finally, we demonstrate the binary classification between PQ and hybrid algorithms.

\textbf{RQ1: Classical vs PQ classification.} 
The datasets consisted of 45,900 samples, with 24,000 classical and 21,900 PQ. Ensemble models, particularly Random Forest and XGBoost, showed superior performance, achieving 98\% overall accuracy in \datasetBNS, with a further increase to 100\% in \datasetANS. These results indicate that classical and PQ algorithms are readily distinguishable based on memory footprint alone. Detailed metrics are provided in Tables \ref{tab:cpq_c} and \ref{tab:cpq}.

\begin{table}[h!]
\centering
\resizebox{\columnwidth}{!}{
\begin{tabular}{|l|c|l|c|c|}
\hline
\textbf{Model} & \textbf{Overall Accuracy} & \textbf{Metric} & \textbf{Classical} & \textbf{Post-Quantum} \\ \hline

\multirow{3}{*}{Logistic Regression} & \multirow{3}{*}{0.99} & Precision & 0.98 & 1.00 \\ \cline{3-5}
 & & Recall & 1.00 & 0.97 \\ \cline{3-5}
 & & F1-Score & 0.99 & 0.99 \\ \hline
\multirow{3}{*}{MLP} & \multirow{3}{*}{1.00} & Precision & 1.00 & 1.00 \\ \cline{3-5}
 & & Recall & 1.00 & 1.00 \\ \cline{3-5}
 & & F1-Score & 1.00 & 1.00 \\ \hline
\multirow{3}{*}{Random Forest} & \multirow{3}{*}{1.00} & Precision & 1.00 & 1.00 \\ \cline{3-5}
 & & Recall & 1.00 & 1.00 \\ \cline{3-5}
 & & F1-Score & 1.00 & 1.00 \\ \hline
\multirow{3}{*}{XGBoost} & \multirow{3}{*}{1.00} & Precision & 1.00 & 1.00 \\ \cline{3-5}
 & & Recall & 1.00 & 1.00 \\ \cline{3-5}
 & & F1-Score & 1.00 & 1.00 \\ \hline

\end{tabular}
}
\caption{Key exchange, classical vs PQ, \datasetA}
\label{tab:cpq_c}
\end{table}

\begin{table}[h!]
\centering
\resizebox{\columnwidth}{!}{
\begin{tabular}{|l|c|l|c|c|}
\hline
\textbf{Model} & \textbf{Overall Accuracy} & \textbf{Metric} & \textbf{Classical} & \textbf{Post-Quantum} \\ \hline
\multirow{3}{*}{Logistic Regression} & \multirow{3}{*}{0.91} & Precision & 0.89 & 0.95 \\ \cline{3-5}
 & & Recall & 0.96 & 0.87 \\ \cline{3-5}
 & & F1-Score & 0.92 & 0.91 \\ \hline
\multirow{3}{*}{MLP} & \multirow{3}{*}{0.97} & Precision & 0.98 & 0.96 \\ \cline{3-5}
 & & Recall & 0.97 & 0.97 \\ \cline{3-5}
 & & F1-Score & 0.97 & 0.97 \\ \hline
\multirow{3}{*}{Random Forest} & \multirow{3}{*}{0.98} & Precision & 0.99 & 0.97 \\ \cline{3-5}
 & & Recall & 0.97 & 0.99 \\ \cline{3-5}
 & & F1-Score & 0.98 & 0.98 \\ \hline
\multirow{3}{*}{XGBoost} & \multirow{3}{*}{0.98} & Precision & 0.99 & 0.97 \\ \cline{3-5}
 & & Recall & 0.97 & 0.99 \\ \cline{3-5}
 & & F1-Score & 0.98 & 0.98 \\ \hline
\end{tabular}
}
\caption{Key exchange, classical vs PQ,  \datasetB}
\label{tab:cpq}
\end{table}

\textbf{RQ2: Multi-class PQ algorithm classification.} 
The datasets comprised 21,000 samples, including 3,000 samples from each PQ algorithm (BIKE, SIKE, CRYSTALS-Kyber, FRODO, Classic-McEliece, HQC, NTRUPrime), with equal representation of all variants. Ensemble learning models, particularly Random Forest and XGBoost, outperformed others, achieving 97\% and 96\% accuracy respectively and strong metrics in identifying individual algorithms within \datasetBNS. In \datasetANS, these models exhibited even higher performance, with 100\% accuracy. These results indicate significant differences in memory usage across post-quantum implementations. Detailed results are presented in Tables \ref{tab:mcpq_c} and \ref{tab:mcpq}.

\begin{table*}[h!]
\centering
\scriptsize % Further reduce the font size
\setlength{\tabcolsep}{2pt} % Further reduce space between columns
\renewcommand{\arraystretch}{1.0} % Adjust space between rows
\begin{adjustbox}{max width=\textwidth}
\begin{tabular}{|p{2.5cm}|p{1cm}|p{1.5cm}|p{1.5cm}|p{1.5cm}|p{1.5cm}|p{1.5cm}|p{1.8cm}|p{1.8cm}|p{1.8cm}|}
\hline
\textbf{Model} & \textbf{Accuracy} & \textbf{Metric} & \textbf{Kyber} & \textbf{Classic Mceliece} & \textbf{HQC} & \textbf{SIKE} & \textbf{BIKE} & \textbf{NTRUPrime} & \textbf{FrodoKEM} \\ \hline
\multirow{3}{*}{Logistic Regression} & \multirow{3}{*}{0.78} & Precision & 0.70 & 0.86 & 0.94 & 1.00 & 0.43 & 0.64 & 0.98 \\ \cline{3-10}
 & & Recall & 0.98 & 1.00 & 0.83 & 1.00 & 0.37 & 0.76 & 0.49 \\ \cline{3-10}
 & & F1-Score & 0.82 & 0.92 & 0.88 & 1.00 & 0.40 & 0.69 & 0.65 \\ \hline
\multirow{3}{*}{XGBoost} & \multirow{3}{*}{1.00} & Precision & 1.00 & 1.00 & 1.00 & 1.00 & 1.00 & 0.99 & 1.00 \\ \cline{3-10}
 & & Recall & 1.00 & 1.00 & 1.00 & 1.00 & 0.99 & 1.00 & 1.00 \\ \cline{3-10}
 & & F1-Score & 1.00 & 1.00 & 1.00 & 1.00 & 1.00 & 1.00 & 1.00 \\ \hline
\multirow{3}{*}{MLP Classifier} & \multirow{3}{*}{0.96} & Precision & 0.98 & 1.00 & 1.00 & 1.00 & 0.95 & 0.81 & 1.00 \\ \cline{3-10}
 & & Recall & 0.96 & 1.00 & 1.00 & 1.00 & 0.75 & 1.00 & 1.00 \\ \cline{3-10}
 & & F1-Score & 0.97 & 1.00 & 1.00 & 1.00 & 0.84 & 0.90 & 1.00 \\ \hline
\multirow{3}{*}{Random Forest Classifier} & \multirow{3}{*}{1.00} & Precision & 1.00 & 1.00 & 1.00 & 1.00 & 1.00 & 1.00 & 1.00 \\ \cline{3-10}
 & & Recall & 1.00 & 1.00 & 1.00 & 1.00 & 1.00 & 1.00 & 1.00 \\ \cline{3-10}
 & & F1-Score & 1.00 & 1.00 & 1.00 & 1.00 & 1.00 & 1.00 & 1.00 \\ \hline
\end{tabular}
\end{adjustbox}
\caption{Key exchange, PQ algorithms,  \datasetA}
\label{tab:mcpq_c}
\end{table*}

\begin{table*}[h!]
\centering
\scriptsize % Further reduce the font size
\setlength{\tabcolsep}{2pt} % Further reduce space between columns
\renewcommand{\arraystretch}{1.0} % Adjust space between rows
\begin{adjustbox}{max width=\textwidth}
\begin{tabular}{|p{2.5cm}|p{1cm}|p{1.5cm}|p{1.5cm}|p{1.5cm}|p{1.5cm}|p{1.5cm}|p{1.8cm}|p{1.8cm}|p{1.8cm}|}
\hline
\textbf{Model} & \textbf{Overall Accuracy} & \textbf{Metric} & \textbf{Kyber} & \textbf{Classic Mceliece} & \textbf{HQC} & \textbf{SIKE} & \textbf{BIKE} & \textbf{NTRUPrime} & \textbf{FrodoKEM} \\ \hline
\multirow{3}{*}{Logistic Regression} & \multirow{3}{*}{0.69} & Precision & 0.61 & 0.69 & 0.42 & 0.98 & 0.72 & 0.97 & 0.49 \\ \cline{3-10}
 & & Recall & 0.94 & 0.56 & 0.47 & 1.00 & 0.45 & 0.99 & 0.42 \\ \cline{3-10}
 & & F1-Score & 0.74 & 0.61 & 0.44 & 0.99 & 0.55 & 0.98 & 0.46 \\ \hline
\multirow{3}{*}{XGBoost} & \multirow{3}{*}{0.96} & Precision & 0.99 & 0.88 & 0.98 & 1.00 & 0.99 & 1.00 & 0.93 \\ \cline{3-10}
 & & Recall & 0.99 & 0.99 & 0.92 & 1.00 & 0.96 & 1.00 & 0.90 \\ \cline{3-10}
 & & F1-Score & 0.99 & 0.93 & 0.95 & 1.00 & 0.97 & 1.00 & 0.92 \\ \hline
\multirow{3}{*}{MLP Classifier} & \multirow{3}{*}{0.88} & Precision & 0.84 & 0.87 & 0.69 & 0.99 & 0.96 & 1.00 & 0.79 \\ \cline{3-10}
 & & Recall & 0.97 & 0.97 & 0.68 & 1.00 & 0.94 & 1.00 & 0.60 \\ \cline{3-10}
 & & F1-Score & 0.90 & 0.92 & 0.69 & 0.99 & 0.95 & 1.00 & 0.68 \\ \hline
\multirow{3}{*}{Random Forest Classifier} & \multirow{3}{*}{0.97} & Precision & 0.99 & 0.88 & 0.98 & 1.00 & 0.99 & 1.00 & 0.96 \\ \cline{3-10}
 & & Recall & 0.99 & 0.99 & 0.93 & 1.00 & 0.97 & 1.00 & 0.91 \\ \cline{3-10}
 & & F1-Score & 0.99 & 0.93 & 0.96 & 1.00 & 0.98 & 1.00 & 0.93 \\ \hline
\end{tabular}
\end{adjustbox}
\caption{Key exchange, PQ algorithms,  \datasetB}
\label{tab:mcpq}
\end{table*}

\textbf{RQ3: \liboqs~ vs \circl~ classification for FRODO and Kyber.} 
FRODO and CRYSTALS-Kyber were the only algorithms shared between the two libraries, resulting in a dataset of 12,000 samples — 3,000 from each algorithm's \liboqs~ and \circl~ implementations, with equal representation across all variants. \circl's implementation includes one variant of FRODO and all variants of CRYSTALS-Kyber, while \liboqs~ supports multiple FRODO variants.
%, which could not be included in the dataset due to the need for consistency. 
Random Forest and XGBoost achieved overall accuracy of 96\% in \datasetBNS, rising to 100\% in \datasetANS. These results demonstrate that libraries are readily distinguishable due to significant differences in cycle counts and memory footprints, influenced by the distinct languages used in their implementations. Detailed metrics are shown in Tables \ref{tab:lib_comparison_c} and \ref{tab:lib_comparison}.

\begin{table}[H]
\centering
\scriptsize
\resizebox{\columnwidth}{!}{
\begin{tabular}{|l|c|l|c|c|c|c|}
\hline
\textbf{Model} & \textbf{Overall Accuracy} & \textbf{Metric} & \textbf{frodokem} & \textbf{frodokem\_circl} & \textbf{kyber} & \textbf{kyber\_circl} \\ \hline

\multirow{3}{*}{Logistic Regression} & \multirow{3}{*}{0.99} 
& Precision & 1.00 & 0.97 & 0.99 & 1.00 \\ \cline{3-7}
& & Recall & 0.97 & 1.00 & 1.00 & 0.99 \\ \cline{3-7}
& & F1-Score & 0.98 & 0.99 & 0.99 & 1.00 \\ \hline

\multirow{3}{*}{XGBoost} & \multirow{3}{*}{1.00} 
& Precision & 1.00 & 1.00 & 1.00 & 1.00 \\ \cline{3-7}
& & Recall & 1.00 & 1.00 & 1.00 & 1.00 \\ \cline{3-7}
& & F1-Score & 1.00 & 1.00 & 1.00 & 1.00 \\ \hline

\multirow{3}{*}{MLP Classifier} & \multirow{3}{*}{1.00} 
& Precision & 1.00 & 1.00 & 1.00 & 1.00 \\ \cline{3-7}
& & Recall & 1.00 & 1.00 & 1.00 & 1.00 \\ \cline{3-7}
& & F1-Score & 1.00 & 1.00 & 1.00 & 1.00 \\ \hline

\multirow{3}{*}{Random Forest Classifier} & \multirow{3}{*}{1.00} 
& Precision & 1.00 & 1.00 & 1.00 & 1.00 \\ \cline{3-7}
& & Recall & 1.00 & 1.00 & 1.00 & 1.00 \\ \cline{3-7}
& & F1-Score & 1.00 & 1.00 & 1.00 & 1.00 \\ \hline
\end{tabular}
}
\caption{Key exchange, \liboqs~vs \circl, \datasetA}
\label{tab:lib_comparison_c}
\end{table}

\textbf{RQ4: PQ vs Hybrid classification for \circl.}
\circl~ was the only library supporting both hybrid and post-quantum algorithms. The datasets comprised 29,998 samples: 15,000 for basic post-quantum algorithms (FRODO, CRYSTALS-Kyber, SIKE) and 14,998 for hybrid algorithms. All models achieved 98\% accuracy for \datasetA and 97\% overall accuracy for \datasetBNS. These results indicate that post-quantum and hybrid implementations are distinguishable, likely due to the increased memory usage and computational cost associated with hybrid approaches. Detailed metrics are provided in Tables \ref{tab:pq_hy_c} and \ref{tab:pq_hy}.

\begin{table}[H]
\centering
\scriptsize
\resizebox{\columnwidth}{!}{
\begin{tabular}{|l|c|l|c|c|c|c|}
\hline
\textbf{Model} & \textbf{Overall Accuracy} & \textbf{Metric} & \textbf{frodokem} & \textbf{frodokem\_circl} & \textbf{kyber} & \textbf{kyber\_circl} \\ \hline

\multirow{3}{*}{Logistic Regression} & \multirow{3}{*}{0.82} & Precision & 0.75 & 0.75 & 0.90 & 0.90 \\ \cline{3-7}
 & & Recall & 0.93 & 0.67 & 0.92 & 0.76 \\ \cline{3-7}
 & & F1-Score & 0.83 & 0.71 & 0.91 & 0.82 \\ \hline
\multirow{3}{*}{XGBoost} & \multirow{3}{*}{0.96} & Precision & 1.00 & 0.87 & 0.99 & 0.98 \\ \cline{3-7}
 & & Recall & 1.00 & 0.99 & 0.98 & 0.86 \\ \cline{3-7}
 & & F1-Score & 1.00 & 0.93 & 0.99 & 0.92 \\ \hline
\multirow{3}{*}{MLP Classifier} & \multirow{3}{*}{0.95} & Precision & 1.00 & 0.87 & 0.99 & 0.95 \\ \cline{3-7}
 & & Recall & 1.00 & 0.98 & 0.96 & 0.86 \\ \cline{3-7}
 & & F1-Score & 1.00 & 0.93 & 0.98 & 0.90 \\ \hline
\multirow{3}{*}{Random Forest Classifier} & \multirow{3}{*}{0.96} & Precision & 1.00 & 0.87 & 0.99 & 0.98 \\ \cline{3-7}
 & & Recall & 1.00 & 0.99 & 0.98 & 0.86 \\ \cline{3-7}
 & & F1-Score & 1.00 & 0.93 & 0.99 & 0.92 \\ \hline
\end{tabular}
}
\caption{Key exchange, \liboqs~vs \circl, \datasetB}
\label{tab:lib_comparison}
\end{table}

\begin{table}[h!]
\centering
\scriptsize
\resizebox{\columnwidth}{!}{
\begin{tabular}{|l|c|c|c|c|}
\hline
\textbf{Model} & \textbf{Overall Accuracy} & \textbf{Metric} & \textbf{Post Quantum} & \textbf{Hybrid} \\ \hline

\multirow{3}{*}{Logistic Regression} & \multirow{3}{*}{0.98} & Precision & 0.96 & 1.00 \\ \cline{3-5} 
 & & Recall & 1.00 & 0.95 \\ \cline{3-5} 
 & & F1-Score & 0.98 & 0.98 \\ \hline
\multirow{3}{*}{MLP} & \multirow{3}{*}{0.98} & Precision & 0.96 & 1.00 \\ \cline{3-5} 
 & & Recall & 1.00 & 0.95 \\ \cline{3-5} 
 & & F1-Score & 0.98 & 0.98 \\ \hline
\multirow{3}{*}{Random Forest} & \multirow{3}{*}{0.98} & Precision & 0.96 & 1.00 \\ \cline{3-5}  
 & & Recall & 1.00 & 0.95 \\ \cline{3-5} 
 & & F1-Score & 0.98 & 0.98 \\ \hline
\multirow{3}{*}{XGBoost} & \multirow{3}{*}{0.98} & Precision & 0.96 & 1.00 \\ \cline{3-5} 
 & & Recall & 1.00 & 0.95 \\ \cline{3-5} 
 & & F1-Score & 0.98 & 0.98 \\ \hline

\end{tabular}
}
\caption{Key exchange, PQ vs Hybrid, \datasetA}
\label{tab:pq_hy_c}
\end{table}

\begin{table}[h!]
\centering
\scriptsize
\resizebox{\columnwidth}{!}{
\begin{tabular}{|l|c|c|c|c|}
\hline
\textbf{Model} & \textbf{Overall Accuracy} & \textbf{Metric} & \textbf{Post Quantum} & \textbf{Hybrid} \\ \hline
Logistic Regression & 0.97 & Precision & 0.94 & 1.00 \\ \cline{3-5} 
 &  & Recall & 1.00 & 0.94 \\ \cline{3-5} 
 &  & F1-Score & 0.97 & 0.97 \\ \hline

MLP & 0.97 & Precision & 0.94 & 1.00 \\ \cline{3-5} 
 &  & Recall & 1.00 & 0.94 \\ \cline{3-5} 
 &  & F1-Score & 0.97 & 0.97 \\ \hline

Random Forest & 0.97 & Precision & 0.94 & 1.00 \\ \cline{3-5} 
 &  & Recall & 1.00 & 0.94 \\ \cline{3-5} 
 &  & F1-Score & 0.97 & 0.97 \\ \hline

XGBoost & 0.97 & Precision & 0.94 & 1.00 \\ \cline{3-5} 
 &  & Recall & 1.00 & 0.94 \\ \cline{3-5} 
 &  & F1-Score & 0.97 & 0.97 \\ \hline

\end{tabular}
}
\caption{Key exchange, PQ vs Hybrid, \datasetB}
\label{tab:pq_hy}
\end{table}

\textbf{Ablation study.}
To validate the robustness of memory usage as a classification metric for key exchange algorithms, we re-ran the models on both datasets, excluding memory-based features and focusing solely on the 12 CPU cycle count features. In \datasetANS, the high-performing models, XGBoost and Random Forest, maintained an accuracy of 90-92\% across classification scenarios. However, in \datasetBNS, accuracy significantly dropped to 69-75\% under similar conditions. These findings suggest that the inclusion of memory-based features in the feature set played a crucial role in preventing substantial accuracy declines between \datasetA and \datasetBNS.

\subsection{Digital Signatures}

In this section, we present our findings for digital signatures following the first three research questions, as there are no hybrid digital signature algorithms.
%across three classification scenarios, which inform our responses to the research questions regarding the fingerprinting of post quantum algorithms in libraries. We begin with the classification between classical and post quantum algorithms, followed by a multi-class classification focused solely on post quantum algorithms. Lastly, we differentiate between the two libraries used in our experiments based on their implemented algorithms.

\textbf{RQ1: Classical vs PQ classification.}
The datasets comprised 53976 samples, with 26976 from post-quantum digital signature schemes and 27000 from classical algorithms. Both \datasetA and \datasetB achieved 100\% accuracy across all models. By reducing the feature set to the four memory-based features used in key exchange classification, we maintained similar accuracy levels. These results indicate that classical and PQ signature schemes are readily distinguishable based on their memory imprints. Detailed metrics are provided in Table \ref{tab:classic_pq_sig}.

\begin{table}[h!]
\centering
\resizebox{\columnwidth}{!}{
\begin{tabular}{|l|c|l|c|c|}
\hline
\textbf{Model} & \textbf{Overall Accuracy} & \textbf{Metric} & \textbf{Classical} & \textbf{Post-Quantum} \\ \hline
\multirow{3}{*}{Random Forest Classifier} & \multirow{3}{*}{1.00} & Precision & 1.00 & 1.00 \\ \cline{3-5} 
 &  & Recall & 1.00 & 1.00 \\ \cline{3-5} 
 &  & F1-Score & 1.00 & 1.00 \\ \hline
\multirow{3}{*}{XGBoost} & \multirow{3}{*}{1.00} & Precision & 1.00 & 1.00 \\ \cline{3-5} 
 &  & Recall & 1.00 & 1.00 \\ \cline{3-5} 
 &  & F1-Score & 1.00 & 1.00 \\ \hline
\multirow{3}{*}{Logistic Regression} & \multirow{3}{*}{1.00} & Precision & 1.00 & 1.00 \\ \cline{3-5} 
 &  & Recall & 1.00 & 1.00 \\ \cline{3-5} 
 &  & F1-Score & 1.00 & 1.00 \\ \hline
\multirow{3}{*}{MLP Classifier} & \multirow{3}{*}{1.00} & Precision & 1.00 & 1.00 \\ \cline{3-5} 
 &  & Recall & 1.00 & 1.00 \\ \cline{3-5} 
 &  & F1-Score & 1.00 & 1.00 \\ \hline
\end{tabular}
}
\caption{Digital signatures, classical vs PQ, both datasets}
\label{tab:classic_pq_sig}
\end{table}

\textbf{RQ2: Multi-class PQ algorithm classification.}
The datasets comprised 48,000 samples, with 12,000 samples from each post-quantum digital signature scheme, ensuring equal representation across all variants. Ensemble learning models, particularly Random Forest and XGBoost, again led in performance, though with lower accuracy compared to key exchange classification. For \datasetBNS, Random Forest and XGBoost achieved 85\% and 86\% overall accuracy, respectively, which improved to 88\% in \datasetANS. These findings suggest that while classifying digital signature schemes may be more complex than key exchange algorithms using the same features, the results remain promising, particularly when computational costs are not influenced by additional load. Detailed metrics are provided in Tables \ref{tab:cpqs_c} and \ref{tab:cpqs}.

\begin{table}[h!]
\scriptsize
\centering
\resizebox{\columnwidth}{!}{
\begin{tabular}{|p{2.3cm}|p{1cm}|p{1cm}|p{1cm}|p{1cm}|p{1cm}|p{1cm}|}
\hline
\textbf{Model} & \textbf{Overall Accuracy} & \textbf{Metric} & \textbf{Dilithium} & \textbf{Falcon} & \textbf{SPHINCS} & \textbf{ML-DSA} \\ \hline

\multirow{3}{*}{Random Forest Classifier} & \multirow{3}{*}{0.88} 
& Precision & 0.78 & 0.95 & 0.98 & 0.81 \\ \cline{3-7}
& & Recall & 0.81 & 0.98 & 0.95 & 0.77 \\ \cline{3-7}
& & F1-Score & 0.80 & 0.97 & 0.96 & 0.79 \\ \hline

\multirow{3}{*}{XGBoost} & \multirow{3}{*}{0.88} 
& Precision & 0.77 & 0.97 & 0.97 & 0.80 \\ \cline{3-7}
& & Recall & 0.81 & 0.97 & 0.97 & 0.76 \\ \cline{3-7}
& & F1-Score & 0.79 & 0.97 & 0.97 & 0.78 \\ \hline

\multirow{3}{*}{Logistic Regression} & \multirow{3}{*}{0.67} 
& Precision & 0.51 & 0.76 & 0.99 & 0.51 \\ \cline{3-7}
& & Recall & 0.54 & 0.99 & 0.68 & 0.47 \\ \cline{3-7}
& & F1-Score & 0.53 & 0.86 & 0.81 & 0.49 \\ \hline

\multirow{3}{*}{MLP Classifier} & \multirow{3}{*}{0.68} 
& Precision & 0.53 & 0.77 & 0.91 & 0.58 \\ \cline{3-7}
& & Recall & 0.77 & 0.93 & 0.71 & 0.32 \\ \cline{3-7}
& & F1-Score & 0.63 & 0.84 & 0.80 & 0.41 \\ \hline
\end{tabular}
}
\caption{Digital signatures, PQ Algorithms, \datasetA}
\label{tab:cpqs_c}
\end{table}

\begin{table}[h!]
\scriptsize
\centering
\resizebox{\columnwidth}{!}{
\begin{tabular}{|p{2.3cm}|p{1cm}|p{1cm}|p{1cm}|p{1cm}|p{1cm}|p{1cm}|}
\hline
\textbf{Model} & \textbf{Overall Accuracy} & \textbf{Metric} & \textbf{Dilithium} & \textbf{Falcon} & \textbf{SPHINCS} & \textbf{ML-DSA} \\ \hline

\multirow{3}{*}{Random Forest} & \multirow{3}{*}{0.85} & Precision & 0.85 & 0.91 & 0.95 & 0.73 \\ \cline{3-7}
 & & Recall & 0.67 & 0.94 & 0.92 & 0.87 \\ \cline{3-7}
 & & F1-Score & 0.75 & 0.92 & 0.93 & 0.79 \\ \hline
\multirow{3}{*}{XGBoost} & \multirow{3}{*}{0.86} & Precision & 0.89 & 0.92 & 0.95 & 0.73 \\ \cline{3-7}
 & & Recall & 0.65 & 0.94 & 0.93 & 0.92 \\ \cline{3-7}
 & & F1-Score & 0.75 & 0.93 & 0.94 & 0.81 \\ \hline
\multirow{3}{*}{Logistic Regression} & \multirow{3}{*}{0.73} & Precision & 0.88 & 0.65 & 0.89 & 0.64 \\ \cline{3-7}
 & & Recall & 0.52 & 0.88 & 0.67 & 0.83 \\ \cline{3-7}
 & & F1-Score & 0.65 & 0.75 & 0.77 & 0.73 \\ \hline
\multirow{3}{*}{MLP Classifier} & \multirow{3}{*}{0.79} & Precision & 0.94 & 0.78 & 0.89 & 0.68 \\ \cline{3-7}
 & & Recall & 0.57 & 0.87 & 0.79 & 0.93 \\ \cline{3-7}
 & & F1-Score & 0.71 & 0.82 & 0.84 & 0.78 \\ \hline
\end{tabular}
}
\caption{Digital signatures, PQ Algorithms, \datasetB}
\label{tab:cpqs}
\end{table}

\textbf{RQ3: \liboqs~vs \circl, for Dilithium.}
The datasets consisted of 18,000 samples, with 9,000 samples each from \liboqs\ and \circl\ implementations. Ensemble learning models, as anticipated, demonstrated superior performance. For \datasetANS, we achieved 100\% accuracy using only 4 features, and 95-96\% accuracy for \datasetBNS, which increased to 100\% when all features were included. Detailed results are presented in Tables \ref{tab:lib_comparison_sc} and \ref{tab:lib_comparison_s}. These findings indicate that the implementations are sufficiently distinct to reveal their corresponding libraries.

\begin{table}[H]
\centering
\scriptsize
\resizebox{\columnwidth}{!}{
\begin{tabular}{|l|c|l|c|c|c|c|}
\hline
\textbf{Model}        & \textbf{Overall Accuracy} & \textbf{Metric} & \textbf{dilithium} & \textbf{dilithium\_circl} \\ \hline
\multirow{3}{*}{Logistic Regression} & \multirow{3}{*}{1.00} & Precision & 1.00 & 1.00 \\ \cline{3-5} 
                                     &                        & Recall    & 1.00 & 1.00 \\ \cline{3-5} 
                                     &                        & F1-Score  & 1.00 & 1.00 \\ \hline
\multirow{3}{*}{MLP}                 & \multirow{3}{*}{1.00} & Precision & 1.00 & 1.00 \\ \cline{3-5} 
                                     &                        & Recall    & 1.00 & 1.00 \\ \cline{3-5} 
                                     &                        & F1-Score  & 1.00 & 1.00 \\ \hline
\multirow{3}{*}{Random Forest}       & \multirow{3}{*}{1.00} & Precision & 1.00 & 1.00 \\ \cline{3-5} 
                                     &                        & Recall    & 1.00 & 1.00 \\ \cline{3-5} 
                                     &                        & F1-Score  & 1.00 & 1.00 \\ \hline
\multirow{3}{*}{XGBoost}             & \multirow{3}{*}{1.00} & Precision & 1.00 & 1.00 \\ \cline{3-5} 
                                     &                        & Recall    & 1.00 & 1.00 \\ \cline{3-5} 
                                     &                        & F1-Score  & 1.00 & 1.00 \\ \hline
\end{tabular}
}
\caption{Digital signatures, \liboqs~vs \circl, \datasetA}
\label{tab:lib_comparison_sc}
\end{table}

\begin{table}[H]
\centering
\scriptsize
\resizebox{\columnwidth}{!}{
\begin{tabular}{|l|c|l|c|c|c|c|}
\hline
\textbf{Model}        & \textbf{Overall Accuracy} & \textbf{Metric} & \textbf{dilithium} & \textbf{dilithium\_circl} \\ \hline
\multirow{3}{*}{Logistic Regression} & \multirow{3}{*}{1.00} & Precision & 0.99 & 1.00 \\ \cline{3-5} 
                                     &                        & Recall    & 1.00 & 0.99 \\ \cline{3-5} 
                                     &                        & F1-Score  & 1.00 & 1.00 \\ \hline
\multirow{3}{*}{XGBoost}             & \multirow{3}{*}{1.00} & Precision & 1.00 & 1.00 \\ \cline{3-5} 
                                     &                        & Recall    & 1.00 & 1.00 \\ \cline{3-5} 
                                     &                        & F1-Score  & 1.00 & 1.00 \\ \hline
\multirow{3}{*}{MLP Classifier}      & \multirow{3}{*}{1.00} & Precision & 1.00 & 1.00 \\ \cline{3-5} 
                                     &                        & Recall    & 1.00 & 1.00 \\ \cline{3-5} 
                                     &                        & F1-Score  & 1.00 & 1.00 \\ \hline
\multirow{3}{*}{Random Forest}       & \multirow{3}{*}{1.00} & Precision & 1.00 & 1.00 \\ \cline{3-5} 
                                     &                        & Recall    & 1.00 & 1.00 \\ \cline{3-5} 
                                     &                        & F1-Score  & 1.00 & 1.00 \\ \hline
\end{tabular}
}
\caption{Digital signatures, \liboqs~vs \circl, \datasetB}
\label{tab:lib_comparison_s}
\end{table}

\subsection{Discussion}

\textbf{Attacker capability to train a model.}
In multi-user/shared environments, such as clusters, an attacker with access to the same machine as the victim can exploit standard system monitoring tools to identify process IDs and access memory usage data, even when the process operates with elevated privileges. Once obtained, this data can be used as test input for a pre-trained machine learning model, which the attacker may have trained locally by running post-quantum and classical algorithms on their own device.

\textbf{Mitigation techniques.}
A potential defense against library classification involves restricting the visibility of memory usage data from processes spawned by one user to others, thereby mitigating unauthorized access to sensitive process metrics. Experimental observations have demonstrated that standard system monitoring tools can be exploited to locate a process’s ID and subsequently access its memory usage file, even when the process operates under elevated privileges. Concealing process identifiers may therefore serve as an effective countermeasure. In addition, memory obfuscation techniques, such as randomized memory allocation—which allocates buffers at random addresses or introduces dummy allocations to maintain a uniform usage profile—and the application of Oblivious RAM schemes to conceal access patterns, offer further protection, albeit with notable performance overheads and primarily as a subject of ongoing research.

\textbf{Limitations.}
Memory-based features offer a more reliable fingerprint than CPU metrics due to inherent differences in memory allocation patterns. PQ algorithms typically require larger buffers for big-integer computations, whereas classical schemes like ECDH utilize significantly smaller memory footprints. These distinctions are readily captured by metrics from /proc/[pid]/status (e.g., VmSize and VmRSS), providing consistent indicators of the underlying algorithm. In contrast, per-core CPU usage metrics from perf often reflect transient activity that is obscured by system-wide noise and background processes, thereby reducing their effectiveness for reliable identification.

\section{Classification of Protocols}

In this section we describe our results on the classification of key exchange algorithms used by a protocol. We seek to answer the following questions:

\begin{itemize} 

\item \textit{RQ1: Given a connection for a particular protocol, can we identify if the key exchange algorithm used by the connection is PQ or classical?} 

\item \textit{RQ2: Given a connection for a protocol, can we identify which PQ key exchange algorithm was used by the connection?} 

\end{itemize}

\subsection{Methodology}
\textbf{Protocols we consider.}
We ran experiments on TLS, SSH, QUIC, Open ID Connect(OIDC) and VPN because of their widespread usage. For our study on TLS, we employed s2n-TLS as the post-quantum (PQ) implementation, while utilizing OpenSSL, Schannel, and Secure Transport API for classical cryptographic implementations. In the context of SSH, we used OpenSSH version 9.2 for PQ implementation and OpenSSH version 8.9 for classical implementation. For QUIC, s2n-QUIC was utilized for PQ implementations. The OIDC protocol was implemented using Post-Quantum-OIDC-OAuth2 for the PQ version. Finally, for VPNs, we used PQ-Crypto-OpenVPN for the PQ implementation and OpenVPN for the classical implementation.

\textbf{Device specifications.} All experiments were conducted across four different devices. The first device was configured with Ubuntu 22.04, running on an Intel Core i5-11400H CPU, featuring 6 physical cores and 12 virtual cores, alongside 16 GB of RAM. The second and third devices were both operating on Windows; the second device was powered by an Intel Core i5-6300U CPU, with 2 physical cores and 4 virtual cores, and the third device utilized an Intel Core i7-8565U CPU, offering 4 physical cores and 8 virtual cores, both equipped with 16 GB of RAM. The final device was running macOS Sonoma, powered by the Apple M2 chip, with 16 GB of RAM.

\textbf{Datasets.}
For local traffic generation and data collection, we utilize packet sniffing tools such as Wireshark and tcpdump. Additionally, we supplement our dataset with publicly available packet captures from sources such as Cloudflare\cite{qacafe_sample_captures} and Wireshark\cite{wireshark_sample_captures}. For local data collection, to emulate real world conditions, we simultaneously run the standard and post quantum implementations so that our packet captures contain not only connections that use classical algorithms but also post quantum. 
 
\subsection{Classification Method}
We aim to identify the key exchange algorithm used for a successful connection between two entities over a network where a successful connection means that the same key exchange algorithm was used on both entities. 
The desirable scenario is to have packets from both entities but if we do not, we also provide a solution to the issue.

We rely just on the clearly transmitted key sizes embedded in the packets to identify and classify the key exchange algorithm used for connection initiation. To achieve this, the packet capture data is passed through a protocol-specific filter. This filter is tailored to the structure of each protocol (e.g., TLS or SSH) and is responsible for selectively extracting only the key exchange–related messages, identified by their specific message types, while excluding all other protocol traffic. The result is a concise list of key exchange algorithms observed within each protocol.

\paragraph{Classification in TLS, OpenVPN, QUIC, OIDC} 
A single approach suffices to identify the key exchange algorithm used by these protocols, as they all rely on some implementation of the TLS protocol to establish secure key exchange. To determine the key exchange method used by a given connection, it is necessary to extract the relevant information from the TLS layer of the packet, which differs between TLS 1.2 and TLS 1.3. Notably, the handshake messages containing the key exchange information are transmitted in the clear—that is, they are not encrypted and are fully observable to any party monitoring the communication.

In a TLS 1.3 handshake, the sequence begins with the client sending a ClientHello message, which includes supported key exchange algorithms, the desired TLS version, and a random value. The server responds with a ServerHello, selecting a key exchange algorithm and providing its own random value. This is followed by EncryptedExtensions for securely exchanging additional information, and if necessary, the server sends its Certificate and CertificateVerify messages to authenticate its identity. The server then concludes its part of the handshake with a Finished message, signaling the completion of the key exchange. In contrast, for a TLS 1.2 connection, the handshake similarly starts with ClientHello and ServerHello messages, but the key exchange details are not included in the ServerHello. Instead, separate server and client key exchange packets follow, with the handshake finalized by Finished messages from both parties. In both TLS 1.3 and TLS 1.2, key sizes are transmitted in the clear during the handshake. The difference in packets containing key exchange information presents us with a challenge that is solved by isolating packets of the protocol containing the server, client hello if the connection is TLS 1.3 and the server, client key exchange packets for a TLS 1.2 connection.

Once we have filtered the correct pair of packets based on the TLS version, we extract the key sizes from the TLS layer of the packets and compare them with the already known key sizes of key exchange algorithms to identify which algorithm was used. Now that we have two algorithms from both packets we compare them to see if the algorithm was part of a successful connection. 

The packets containing key exchange information from both the client and server may occasionally go undetected by our approach or may not be captured within the observation period. If we have just the client hello, our approach is unable to detect the algorithm. If we have instead the server hello/server key exchange/client key exchange packet our method is unable to compare the algorithm for both packets and may provide multiple suggestions for the key exchange algorithm which include the correct one. Manual analysis of the client side packet gives the correct algorithm.

%\cnr{Tushin, add something about the number of messages in the handshake} \tushin{There is no noticeable distinction in the number of messages for TLS. I will check again tomorrow}
%\cnr{that does not make sense, when TLS does diffie helman both sides send their contribution. With pq it seems to me that one party selects the key and just sends it. Do both client and server send a key to each other?}

\paragraph{Classification in SSH}
This is mostly similar to the previous approach. Since SSH messages do not have a TLS layer, the key exchange messages are instead sent via the ssh layer. So, in this case we isolate the Client ECDH Key Exchange Init and the Server ECDH Key Exchange Init packets and then extract the key size from the ssh layer. Then we proceed as in the previous approach.

\subsection{Results}

\textbf{RQ1: Classical vs PQ protocols.}
For SSH connections, on Ubuntu, MacOS and Windows we successfully identify all classical and post quantum connections. Since OpenSSH(version$\geq$9.0) supports only 1 hybrid algorithm by default, all post quantum connections used that key exchange. As for TLS connections we achieved complete success for connections between devices running Ubuntu and MacOS.  Our classification approach is able to detect and identify key exchanges for most QUIC connections. All TLS connections for key exchange within entities of OIDC such as the relying party, user agent and OIDC provider were also successfully identified by our approach. Based on our findings we can say that classical and post quantum algorithms are easily distinguishable from their key sizes as the post-quantum public keys and cipher texts tend to be exponentially larger than their classical counterparts.

\textbf{RQ2: Multi-class PQ algorithm classification.}
Our findings also reveal that the PQ algorithms themselves are pretty distinct when it comes to key sizes. Even when hybrid algorithms are used by protocols, the keys sizes seldom collide with that of post quantum algorithms and our approach is successful in identifying them.

\subsection{Discussion}

\textbf{Attacker capability to conduct the classification.}
An adversary on the same network as the victim can intercept handshake packets exchanged between the victim and external servers, allowing them to determine whether the connection employs post-quantum key exchange. Since key sizes and public keys are transmitted in plaintext within the ClientHello and ServerHello messages, the adversary can analyze these parameters to infer the cryptographic scheme in use. This exposure enables the identification of connections relying on vulnerable key exchange algorithms, potentially allowing the adversary to exploit weaknesses and compromise security.

\textbf{Defenses. }
A potential mitigation for protocol identification is to ensure that key shares used in deriving a shared secret are not transmitted in plaintext over the network. Although the Encrypted Client Hello (ECH) effectively encrypts the client’s key share—thereby preventing its exposure—this approach does not extend to the server, whose key share remains transmitted in clear during the Server Hello, thereby presenting a vulnerability that could facilitate identification.

\textbf{Limitations. }
Our approach misses some TLS connections from Windows machines because of custom headers. Although our approach is also able to handle fragmented packets in most cases, fragmentation of client and/or server hello packets in TLS 1.2 for OpenVPN connections fails to be recognized by it. 
%We also lack the support for TLS 1.0 as its usage is not recommended and supported by the current systems.

\subsection{Integration of PQ digital signatures in secure protocols}

Current PQ secure protocols has been limited to replacing the DH key exchange with a KEM scheme. Complete solutions should also
consider the integration of digital signatures. 

In TLS, the use of larger PQ signatures along with PQ key exchange during the handshake will overshoot the typical Ethernet MTU of 1,500 bytes after accounting for protocol overheads, resulting in packet fragmentation. For example, if NIST-standardized Dilithium scheme is employed, the signature sizes exceed the MTU  necessitating packet fragmentation. Specifically, a Dilithium 2 signature (2420 bytes) will need 2 packets, while a Dilithium 3 signature (3293 bytes) will require 3 packets to be transmitted. Fragmentation not only increases latency due to reassembly and retransmission but also enhances fingerprintability, as the distinct size and fragmentation patterns of PQ handshake packets can expose cryptographic traffic to adversaries. These scalability, performance, and security challenges underscore the need for significant advancements in router hardware, network bandwidth, and cryptographic efficiency before widespread adoption of PQ signature schemes in such protocols. 

Because there are no current secure protocols that have replaced digital signatures, we investigated the possibilty of distinguishing versions using PQ solutions, by analyzing digital signatures when the only information available is latency, where the additional cost introduced by PQ schemes is going to be computational cost. 
Given the superior performance of XGBoost compared to other models we experimented with, it was selected for further training using both \textbf{\datasetA} and \textbf{\datasetB}. The model was trained without memory-related parameters, focusing solely on CPU cycle data, to determine how effectively multi-class classification of digital signature algorithms could be achieved based on this subset of features. The dataset consisted of 80,000 samples, with 20,000 samples from each post-quantum digital signature scheme, ensuring equal representation across all variants. The accuracy remained consistent, ranging between 84\% and 86\%.

\begin{comment}
The adoption of post-quantum (PQ) signature schemes, such as Dilithium—a NIST-standardized algorithm—introduces significant challenges in protocols like Border Gateway Protocol (BGP) and Transport Layer Security (TLS) due to the substantial size of their signatures. While traditional algorithms like RSA-2048 and ECDSA-256 have compact signature sizes (256 bytes and 64 bytes, respectively), Dilithium signatures range from ranges from 2,420 bytes (Dilithium-2) to 4,595 bytes (Dilithium-5) depending on the security level. In BGP, incorporating such large signatures into global routing tables containing millions of entries could dramatically increase memory requirements and bandwidth usage, potentially by an order of magnitude. The computational overhead of verifying these signatures further risks delaying route propagation making such incorporation infeasible without substantial advances in router hardware, network bandwidth, and cryptographic efficiency.
\end{comment}
\section{Classifying SNARK libraries}
In this section, we describe the classification of SNARK generation libraries. %We first describe our methodology, followed by the algorithm and results. 
We seek to answer the following question:

\begin{itemize} 

\item \textit{RQ1: Is it possible to distinguish between classical and PQ SNARK generation library implementations?}

\end{itemize}

\subsection{Methodology}

\textbf{Libraries we consider.}
We investigate two SNARK libraries: lattice-zkSNARK \cite{ISW21}, which integrates post-quantum cryptographic techniques within the libsnark framework \cite{libsnark}, and pysnark \cite{pysnark}, which relies on the classical libsnark backend. For a rigorous comparative analysis, both libraries were evaluated by generating SNARKs for a standardized program.

\textbf{Device specifications.}
Experiments were conducted on a single device equipped with
Ubuntu 22.04, powered by an Intel Core i5-11400H CPU, which
features 6 physical cores and 12 virtual cores, complemented by 16 GB of RAM.

\textbf{Datasets.}
CPU cycle count and memory usage during SNARK generation and verification were used as classification features. CPU cycle counts were obtained using the perf command, and memory usage was measured from /proc/[pid]/status over 1000 experimental runs. Two datasets were generated: \textbf{\datasetAsnark}, which minimized background processes, and \textbf{\datasetBsnark}, which incorporated CPU-intensive tasks such as matrix multiplications, prime number calculations, large list sorting, and hash computations. While these additional tasks influenced computation time, they did not affect memory usage for the evaluated processes. Owing to limited data availability, the final dataset used in this study comprised a combination of \datasetAsnark and \datasetBsnark.

\subsection{Classification Method}
\begin{figure}[h!]
  \centering
  \includegraphics[width=\linewidth]{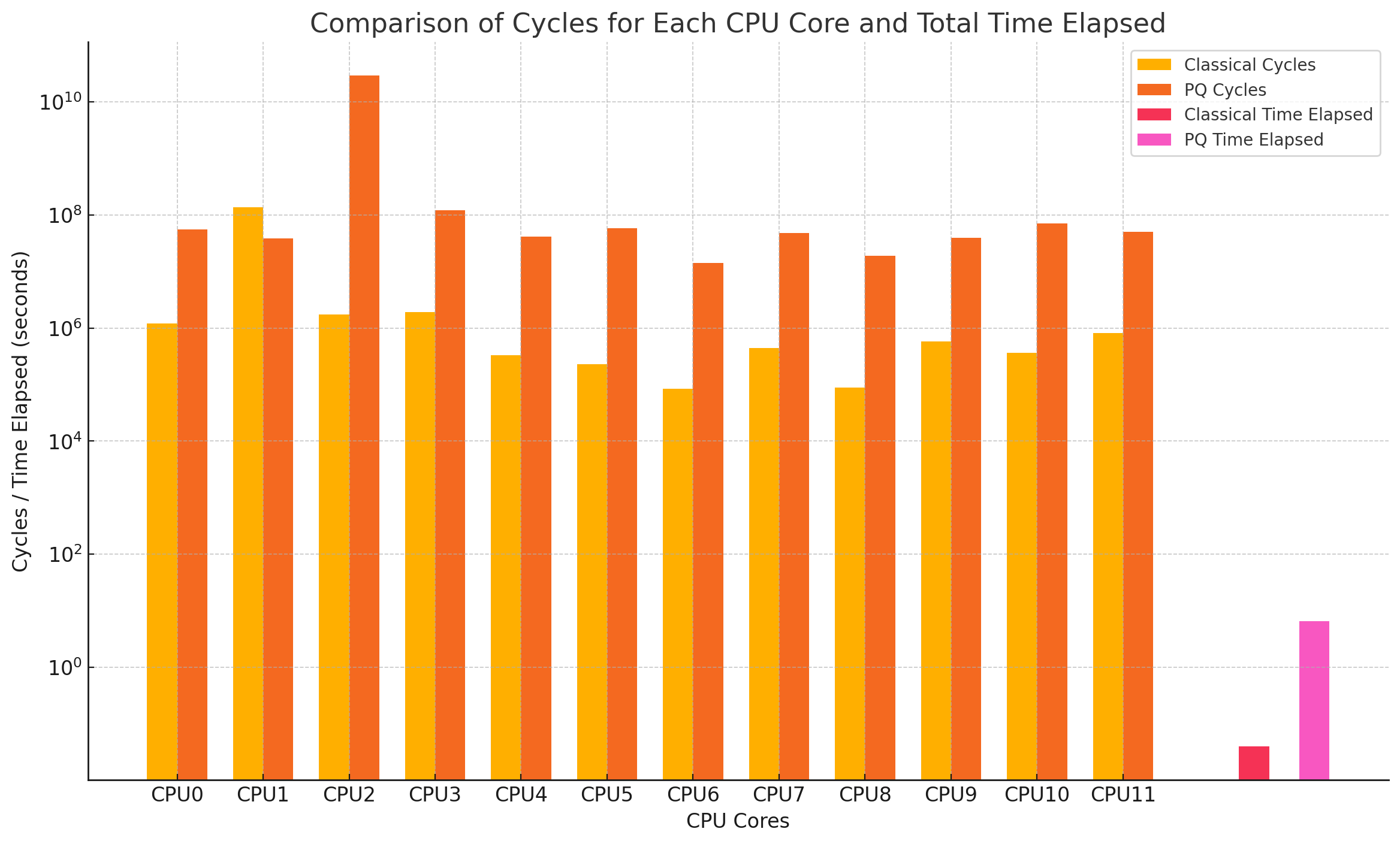} 
  \caption{SNARK libraries, \pysnark ~vs \lattice, CPU cycle comparison}
  \label{fig:cycle_snarks}
\end{figure}

\begin{figure}[h!]
  \centering
  \includegraphics[width=\linewidth]{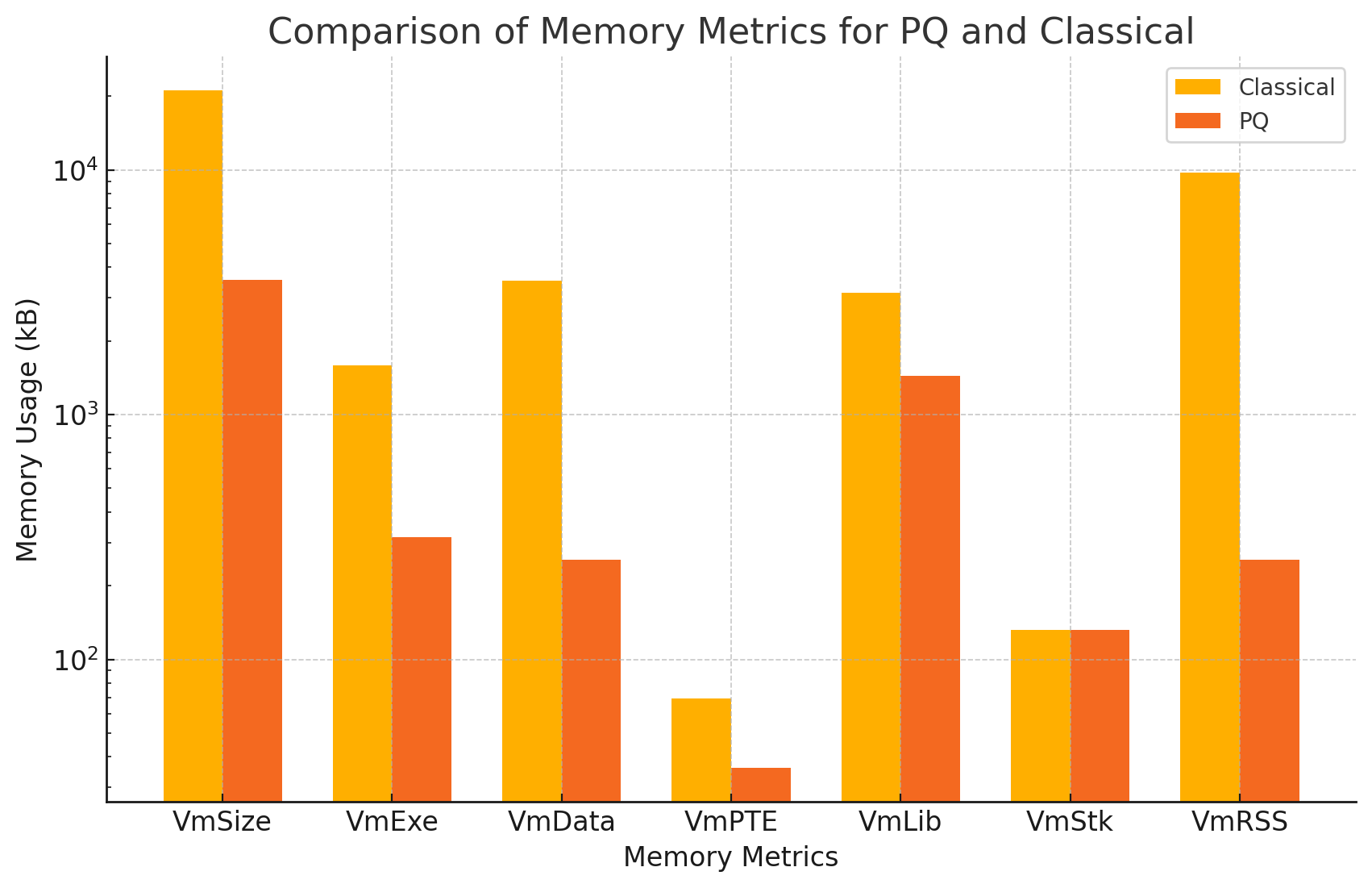} 
  \caption{SNARK libraries, \pysnark ~vs \lattice, Memory usage comparison}
  \label{fig:mem_snarks}
\end{figure}

Our initial analysis reveals significant differences between the two libraries in terms of computational cycles and memory usage, enabling clear library-specific classification. For example, as seen from Fig \ref{fig:cycle_snarks} and \ref{fig:mem_snarks}, the PQ lattice-zkSNARK have substantially higher execution times compared to the classical pysnark, reflecting the additional cryptographic complexity inherent in post-quantum systems. Moreover, the distinct cycle distributions across CPU cores and unique memory footprints associated with each library further reinforce their identifiable characteristics. These findings underline the potential for accurate differentiation of SNARK implementations, offering valuable insights for security analysis and optimization in both post-quantum and classical SNARK contexts.

Based on these results, we formulate the problem as a machine learning binary classification, with classes identified as classical and post-quantum. We start the classification task with 19 features, of which 12 features represent the cycles used by each core of the CPU and remaining 7 features represent a process’s memory usage, including total virtual memory, physical memory in use, data segment, stack, executable
code, shared libraries, and page table entries. We narrow down our feature set to 2 CPU cycle based features using the Chi-square statistical test as the scoring function to get the top features by importance. Finally, we use an 80-20 train-test split of the dataset.

\subsection{Results}

The dataset comprised of 4000 samples, evenly split between \pysnark~ and \lattice~. The XGBoost model achieved 100\% classification accuracy. The results are presented in Table \ref{tab:classification_snarks}. This level of accuracy can be extrapolated to larger datasets, given the distinctive and consistent patterns observed in the computational and memory usage metrics.

\begin{table}[h!]
\centering
\scriptsize
\resizebox{\columnwidth}{!}{
\begin{tabular}{|l|l|l|l|l|}
\hline
\textbf{Model}   & \textbf{Overall Accuracy} & \textbf{Metric} & \textbf{pysnark (classical)} & \textbf{lattice\_zksnark (post-quantum)} \\ \hline
\multirow{3}{*}{XGBoost}          & \multirow{3}{*}{1.00}                     & Precision       & 1.00                         & 1.00                                   \\ \cline{3-5}
                 &                           & Recall          & 1.00                         & 1.00                                   \\ \cline{3-5}
                 &                           & F1-Score        & 1.00                         & 1.00                                   \\ \hline
\end{tabular}
}
\caption{SNARK libraries, \pysnark ~vs \lattice, \datasetAsnarkNS \ \datasetBsnarkNS}
\label{tab:classification_snarks}
\end{table}

\section{Case Studies}

In this section we describe how classification of post-quantum libraries and protocols can benefit real world applications.

\subsection{Quartz Integration}

As part of the real-world transition to PQC~\cite{nist-pqc-details}
enterprises must be able to assess their PQC-readiness.
To support these efforts, Cisco has released as open source a research prototype called "Quartz" - Quantum Risk and Threat Analyzer~\cite{quartz}. Quartz supports static and dynamic scanning of several protocols and processes, analyzes and identifies quantum-vulnerable cryptography protocols and libraries being used, and carries out risk analysis that can help in remediation of such vulnerabilities using quantum-resistant algorithms and implementations such as PQC standards.  It scans and analyzes network communications, cloud applications, databases, operating systems, file systems. Quartz also analyzes source code, SQL queries, and cloud account activities.

%Our fingerprinting approaches complement Quartz. 
Quartz  relies on communication meta-data and payload if it can access, the configurations and code/scripts in order to determine the type of cryptography protocols, but it lacks the ability to analyze performance metrics and classification capabilities as proposed in this paper.  Our classification techniques can be used by a system like Quartz in order to be able to identify quantum-vulnerable cryptography protocols and libraries across the communication and computing stack in a black-box and/or white-box manner. 
%Quartz can thus be able to provide detailed algorithmic level details to the vulnerability analyzer, the risk analysis engine and assess the risk score in an improved manner.  In turn remediation of quantum vulnerabilities  can benefit significantly with such an augmented  Quartz system.

\textbf{Integration with Quartz.}
We enhanced the Quartz system by incorporating advanced features for post-quantum connection detection in TLS and the classification of cryptographic key exchange and digital signature algorithms based on CPU cycle and memory usage. Furthermore, we have developed and integrated dedicated API endpoints into the Quartz system, facilitating seamless backend execution of the relevant scripts.  We will make our changes available in the main Quartz repository to promote transparency and accessibility. We provide an anonymous repository with a cloned version of Quartz and our integrated features, available at \url{https://anonymous.4open.science/r/Fingerprint-809D/}.

The TLS connection detection feature is designed to identify connections employing PQ cryptographic algorithms for key exchange. This functionality is enabled by submitting a .pcapng file through a POST request to the /classify API endpoint. The system processes the file and outputs the IP addresses utilizing PQ connections along with the probable key exchange algorithm(s). The feature supports efficient analysis of network traffic and is particularly suited for monitoring environments transitioning to PQ cryptographic standards.

The second major enhancement focuses on classifying cryptographic algorithms based on their resource usage. For this task, the system leverages models trained on both \textbf{\datasetA} and \textbf{\datasetB} to ensure resilience in diverse operational contexts. To classify key exchange algorithms, users submit a .csv file containing CPU cycle and memory usage metrics to the /classifyKex endpoint. Similarly, digital signature algorithms can be classified by providing a .csv file to the /classifySig endpoint. The outputs from these endpoints consist of the predicted algorithm associated with each entry in the dataset, enabling detailed profiling of cryptographic operations.

\subsection{Identifying Post Quantum TLS Connections on domains from Tranco}

To identify servers on the internet using PQC keys for TLS connections, we conducted a targeted probe of 1 million domains selected from the Tranco dataset \cite{LePochat2019tranco}. This effort yielded 160,023 connection attempts to these domains, from which 4,988 unique IP addresses were identified as potentially supporting PQ keys. It is worth noting that not all domains could be probed due to DNS resolution failures, impacting the total connection attempts.

Occasionally, our probing process generated multiple algorithmic suggestions per connection, including both classical and PQ methods. For instance, the IP associated with the domain 1-800-FLOWERS.com suggested the use of either ECDH-P384 (classical) or Classic-McEliece-348864 (PQ) key, but manual validation confirmed the use of ECDH-P384. While these minor discrepancies introduce a small margin of error, we maintain that a slight overestimation of PQC presence is preferable to underdetection in this exploratory context. Table \ref{tab:ips} presents the top 10 organizations ranked by the number of IP addresses associated with organisations.

\begin{table}[h!]
\centering
\resizebox{\columnwidth}{!}{%
\begin{tabular}{|>{\raggedright}p{5cm}|>{\centering\arraybackslash}p{3cm}|}
\hline
\textbf{Organization} & \textbf{Number of IPs} \\
\hline
Cloudflare, Inc. & 4083 \\
Google LLC & 170 \\
Amazon Technologies Inc. & 58 \\
Microsoft Corporation & 51 \\
RIPE Network Coordination Centre & 49 \\
DigitalOcean, LLC & 48 \\
Akamai Technologies, Inc. & 25 \\
Shopify, Inc. & 18 \\
Leaseweb USA, Inc. & 10 \\
OVH Hosting, Inc. & 10 \\
\hline
\end{tabular}%
}
\caption{Top 10 Organizations by Number of IPs}
\label{tab:ips}
\end{table}

\section{Related Work}

In this section, we review the related literature. We begin by examining side-channel attacks targeting traditional cryptographic algorithms, followed by an analysis of fingerprinting techniques applied to protocols through encrypted traffic analysis. Next, we examine the benchmarking of cryptographic protocols and algorithms to assess their performance and resilience in various environments.

\textbf{Side-channel attacks.} The prevalence of side-channel attacks on classical cryptographic algorithms, such as RSA, AES, and ECC, provides a strong precedent for the potential occurrence of similar attacks on PQ algorithms. Walter et al. \cite{walter2004} revealed that side-channel attacks like simple power analysis can potentially reveal the secret key through power consumption variations during the execution of ECC point operations. Kocher et al. \cite{kocher1999} showed that Differential Power Analysis can be used to extract AES keys by analyzing power consumption patterns during encryption operations. Genkin et al. \cite{rsaAcoustic} presents an acoustic side-channel attack that can extract RSA keys by analyzing the sound produced by the computer during cryptographic operations. Brumley et al. \cite{brumley2005remote} showed that they can extract private keys from an OpenSSL-based web server running on a machine in the local network by timing side-channel attacks against OpenSSL. Wang et al. \cite{wang2023dvfs} further demonstrates that dynamic voltage and frequency scaling (DVFS) can cause data-dependent power consumption to influence CPU frequency, creating timing variations even in “constant-time” implementations and enabling remote key recovery and cross-component leakage. \textsc{Crystals}-Kyber implementations have also been broken by Dubrova et al.~\cite{dubrova2023breaking} (after it was selected as a finalist and before the standardization).

\textbf{Traffic identification.}
Wright et al.  \cite{wright2006inferring} shows a novel approach to protocol identification in encrypted traffic by utilizing observable features such as packet timing, size, and direction, achieving high accuracy without relying on packet contents or host information. Wang et al. \cite{1dcnn} proposes an end-to-end encrypted traffic classification method with only a one-dimensional convolution neural network. Xue et al. \cite{xue2022} demonstrated how the network flow of OpenVPN could be fingerprinted by identifying fixed patterns within OpenVPN traffic. Similarly, Panchenko et al. \cite{torf} presented an attack on Tor, revealing that traffic analysis could be used to identify visited websites, despite the use of encryption, by exploiting unique traffic patterns generated by different websites.

\textbf{Benchmarking PQ protocols.}
Paquin et al. \cite{PQCrypto:PaqSteTam20} introduces a framework for benchmarking PQ cryptographic algorithms within TLS, specifically analyzing the impact of latency and packet loss on hybrid key exchanges and digital signatures.

\section{Conclusion}\label{sec:conclusion}

In this work, we demonstrate the feasibility of distinguishing between post-quantum and classical algorithms based on their implementation within cryptographic libraries, SNARK generating libraries and widely used protocols. We classified and identified library implementations for both post-quantum and classical primitives and SNARKs. The classification is done by analyzing the CPU cycle counts and memory footprints associated with these algorithms, utilizing ensemble learning models that deliver high accuracy in differentiation. We extended our analysis to the identification of key exchange algorithms employed within various connections, specifically focusing on post-quantum implementations of protocols such as TLS, SSH, OpenID Connect, QUIC, and VPN. To achieve this, we meticulously analyze both self-generated and publicly available packet captures. By filtering for key exchange packets, we are able to accurately discern the specific key exchange algorithms being utilized. Finally, we integrate our classification methods with Quartz and also identify domains in Tranco which use post qauntum TLS handshakes.

%This work contributes to the broader field of cryptography by offering a novel method for the detection and differentiation of cryptographic algorithms, particularly in the context of the emerging post-quantum cryptographic landscape. Our findings provide a foundational approach for the automated recognition of cryptographic protocols and their underlying algorithms, which is essential for enhancing security measures in the face of advancing computational threats, as well as towards hardening the algorithms and implementations against fingerprinting attacks.

\bibliographystyle{IEEEtran}
\bibliography{bib/main}

\appendices

\end{document}